\documentclass[aps,prd,twocolumn,superscriptaddress,a4paper,nofootinbib]{revtex4}
\usepackage{xcolor}
\usepackage{tabularx}
\usepackage{multirow}

\usepackage{graphics} 
\usepackage{graphicx}
\usepackage{amsmath} 
\usepackage{float}
\usepackage{natbib}
\usepackage{hyperref}
\begin{document}
\title{\LARGE \bf Constraints on dark photons and axion-like particles from SuperCDMS Soudan}
\author{T.~Aralis} \affiliation{Division of Physics, Mathematics, \& Astronomy, California Institute of Technology, Pasadena, CA 91125, USA}
\author{T.~Aramaki} \affiliation{SLAC National Accelerator Laboratory/Kavli Institute for Particle Astrophysics and Cosmology, Menlo Park, CA 94025, USA}
\author{I.J.~Arnquist} \affiliation{Pacific Northwest National Laboratory, Richland, WA 99352, USA}
\author{E.~Azadbakht} \affiliation{Department of Physics and Astronomy, and the Mitchell Institute for Fundamental Physics and Astronomy, Texas A\&M University, College Station, TX 77843, USA}
\author{W.~Baker} \affiliation{Department of Physics and Astronomy, and the Mitchell Institute for Fundamental Physics and Astronomy, Texas A\&M University, College Station, TX 77843, USA}
\author{S.~Banik} \affiliation{School of Physical Sciences, National Institute of Science Education and Research, HBNI, Jatni - 752050, India}
\author{D.~Barker} \affiliation{School of Physics \& Astronomy, University of Minnesota, Minneapolis, MN 55455, USA}
\author{C.~Bathurst} \affiliation{Department of Physics, University of Florida, Gainesville, FL 32611, USA}
\author{D.A.~Bauer} \affiliation{Fermi National Accelerator Laboratory, Batavia, IL 60510, USA}
\author{L.V.S.~Bezerra} \affiliation{Department of Physics \& Astronomy, University of British Columbia, Vancouver, BC V6T 1Z1, Canada} \affiliation{TRIUMF, Vancouver, BC V6T 2A3, Canada} 
% ...
\collaboration{SuperCDMS Collaboration}

\begin{abstract}
We present an analysis of electron recoils in cryogenic germanium detectors operated during the SuperCDMS Soudan experiment. The data are used to set new constraints on the axioelectric coupling of axion-like particles and the kinetic mixing parameter of dark photons, assuming the respective species constitutes all of the galactic dark matter. This study covers the mass range from $40$\,eV/$c^2$ to $500$\,keV/$c^2$ for both candidates, excluding previously untested parameter space for masses below $\sim$1\,keV/$c^2$. For the kinetic mixing of dark photons, values below 10$^{-15}$ are reached for particle masses around 100\,eV/$c^2$; for the axio-electric coupling of axion-like particles, values below 10$^{-12}$ are reached for particles with masses in the range of a few-hundred eV/$c^2$.
\end{abstract}
\maketitle

\section{Introduction}
Many astrophysical and cosmological observations support the existence of dark matter, which constitutes more than 80$\,$\% of the matter in the universe \cite{Ade:2015xua, Clowe:2006eq}. While the current Standard Model of particle physics (SM) does not describe dark matter, a number of theoretical extensions predict new particles that are viable dark matter candidates. For the past three decades Weakly Interacting Massive Particles (WIMPs) have been the primary candidate of interest \cite{Hochberg:2016sqx}. Recently however, low-mass candidates ($\mathcal{O}$(keV/$c^2)$ or below) such as axions or axion-like particles (ALPs) and dark photons have gained traction \cite{Hochberg:2016sqx}; these particles occur naturally in many proposed models, and have been invoked to explain various experimental and observational anomalies \cite{Peccei:2006as, Svrcek:2006yi, Raggi:2015yfk}. Both axions/ALPs and dark photons may be produced in the early universe and thus may constitute a significant fraction (if not all) of the dark matter in the universe \cite{Sikivie:2006ni, Nelson:2011sf, Co:2018lka, Long:2019lwl}. In theories that include ALPs or dark photons, a coupling between the new particle and a SM particle may lead to a process by which the new particle is absorbed by an atom and an electron is ejected, carrying away the excess energy \cite{Bloch:2016sjj}. This process is analogous to the photoelectric effect and is henceforth referred to as dark absorption. As galactic dark matter is non-relativistic, the observable signature from relic ALPs or dark photons would be a peak in the recoil spectrum at the rest mass energy of the particle.

The Super Cryogenic Dark Matter Search (SuperCDMS) experiment \cite{PhysRevLett.120.061802}, located in the Soudan Underground Laboratory in Northern Minnesota, used cryogenic germanium detectors to search for signals produced by dark matter. Particle interactions within the detector produce phonon and ionization signals, which are measured using superconducting transition edge sensors (TES) and charge electrodes, respectively. The ratio of the two signals is different for interactions with nuclei or electrons, providing an efficient discrimination tool \cite{PhysRevLett.120.061802}. The interleaved layout of the charge and phonon sensors allows for the further discrimination of events near the electrodes, where they can suffer from reduced charge collection \cite{Agnese:2013}, giving rise to the name {\em interleaved Z-sensitive Ionization and Phonon} (iZIP) detectors. In the CDMS Low Ionization Threshold Experiment (CDMSlite), a much higher bias voltage is applied across the detector to make use of the Neganov-Trofimov-Luke (NTL) effect \cite{Neganov,Luke}, in which additional phonons are created in proportion to the number of drifting charges and the magnitude of the bias voltage. This effect leads to a sensitivity to considerably lower-energy interactions and thus lighter dark matter particles. However, the discrimination between electron and nuclear recoils is lost. Additionally, the increased amplification causes saturation effects to occur at lower energies than in the regular iZIP operating mode, lowering the upper end of the usable energy range.

Traditionally, SuperCDMS dark matter analyses have focused on searches for WIMPs \cite{steigman} scattering off detector nuclei. In this paper, electron recoil data from SuperCDMS Soudan are analyzed to constrain the parameters that describe the absorption of ALPs and dark photons. The expected signal from this process is a peak in the energy spectrum at the energy corresponding to the mass of the dark matter particle, with a width that is given by the resolution of the detector at that energy. For this analysis, we do not model or subtract the background; thus, only upper limits on the rates of dark absorption can be set. Conservative limits are placed on the coupling of ALPs to electrons and the kinetic mixing between the dark and SM photons for particle masses between 40\,eV/$c^2$ and 500\,keV/$c^2$, where CDMSlite data cover the lower and iZIP data the higher masses.

An overview of the two dark matter candidate models, the expected signals in the detector, and the assumptions used to determine limits on the couplings are given in Sec.\,\ref{sec:theory}. In Sec.\,\ref{sec:cdms} the experimental setup of SuperCDMS Soudan is described. Section\,\ref{sec:analysis} describes the key steps in the data analysis and the method used to derive limits on the coupling parameters, with the results presented in Sec.\,\ref{sec:results} and discussed in the concluding Sec.\,\ref{sec:conclusion}.


\section{\label{sec:theory}Theory Overview}


\subsection{\label{sec:ALPs}Axions and ALPs}

The axion was originally proposed to account for the apparent fine-tuning associated with the lack of charge-parity (CP) violation in the strong interaction \cite{Peccei:2006as}. This phenomenon could be naturally explained with a new, spontaneously broken, approximate global $U(1)$ symmetry; the axion would be the pseudo-Goldstone boson associated with this spontaneous symmetry breaking \cite{Peccei:2006as}. Other spontaneously broken global symmetries appear in many extensions to the SM (such as string theories \cite{Svrcek:2006yi,Marsh:2017hbv}), giving rise to axion-like particles (ALPs). Axions and ALPs would both feature an effective coupling to the SM photon \cite{diCortona:2015ldu, Marsh:2017hbv} and to electrons \cite{Derevianko:2010kz}. However, the mass of the canonical axion (associated with the strong CP problem) is limited to be less than $\sim$10$^{-2}$\,eV/$c^2$ by the duration of
the neutrino signal from SN 1987A \cite{Raffelt:1995ym} and the cooling of neutron stars \cite{PhysRevD.93.065044}, meaning it is too light to excite an electron in a SuperCDMS detector through the absorption process. These constraints are dependent on the required coupling of axions to nucleons; since ALPs don't require such a coupling, the constraints do not apply. Therefore, only ALPs (and not axions) will henceforth be discussed.

\subsection{\label{sec:DP}Dark Photons}
Dark photons are hypothetical vector bosons that would mediate a new dark force in models with a new local $U(1)$ symmetry. Kinetic mixing with the SM photon \cite{Holdom:1985ag} enables the dark photon to interact with electrically charged particles. Although the dark photon does not necessarily need to be accompanied by other particles outside of the SM, it is usually invoked as part of a dark sector, where it serves as a mediator between the dark sector and the SM. The kinetic mixing is then generated by loop-level interactions of much heavier dark particles, charged under both SM electromagnetism and the new $U(1)$ charge. There is no general constraint on the mass of dark photons in these models. However, in order to be a viable dark matter candidate its mass must be less than twice the electron mass; otherwise, its relic abundance would be depleted by decays into electron-positron pairs \cite{Bloch:2016sjj}.

\subsection{\label{sec:ratecalculation}Dark Absorption}


The absorption cross sections for ALPs and dark photons relate the observed rate to physical quantities relevant to the particular candidate: namely the axioelectric coupling in the case of ALPs, and the kinetic mixing between the dark and SM photons. These cross sections depend on the photoelectric cross section of the target atom (in this case Ge).

We assume the dark matter candidate is non-relativistic and constitutes all of the galactic dark matter. The galactic dark matter flux $\Phi = \rho_{\chi} v_{\chi}/m_{\chi}$ depends on the local dark matter density $\rho_{\chi}$, mass $m_{\chi}$, and velocity $v_{\chi}$. Throughout this work a local dark matter density of $\rho_{\chi} = 0.3$\,GeV\,/($c^{2} \, \mathrm{cm}^{3})$ \cite{Arisaka:2012pb} is assumed. As we will see, the cross sections in question are inversely proportional to the dark matter particle's velocity and therefore cancel the velocity dependence in the flux, making this search velocity independent. 

It should be noted that the predicted dark matter signal rates are an approximation under the assumption that the dark matter Compton wavelength is much larger than the size of the atom \cite{Pospelov:2015}. This is a common assumption in present literature for the calculation of sensitivity limits \cite{Pospelov:2015,Armengaud:2018cuy, Abe:2018owy,Abgrall:2016tnn, Liu:2016osd} and limit projections \cite{Hochberg:2016sqx,Bloch:2016sjj}. This assumption requires corrections at dark matter masses above $\sim$10--100~keV/$c^2$ \cite{Pospelov:2015,Dzuba:2010cw}. More accurate rate predictions taking multi-body and relativistic effects into account require dedicated calculations of dark matter absorption by the atom \cite{DAmodels}. Such calculations do not yet exist for dark photons. For ALP absorption in Ge they exist only for masses up to 100\,keV/$c^2$ \cite{Derevianko:2010kz}, depending on the target material, which does not cover the full energy range for our search. For this reason, and for consistency with existing publications, the presented analysis applies this imperfect assumption.

\subsubsection{Axioelectric Effect}
The effective ALP-electron interaction is quantified by the axioelectric coupling $g_{ae}$. Dark absorption of an ALP via the axioelectric effect would eject a bound electron from an atom. The expected cross section $\sigma_a$ is proportional to the photoelectric cross section $\sigma_{pe}$ \cite{Pospelov:2008jk, Fu:2017lfc} of the target

\begin{equation}
    \sigma_a(E_a) = \sigma_{pe}(E_a) \frac{g_{ae}^2}{\beta_a} \frac{3 E_a^2}{16 \pi \,\alpha\, m_e^2 c^4} \left( 1 - \frac{\beta_a^{2/3}}{3} \right),
\end{equation}

\noindent where $E_a$ is the ALP's total energy, $\beta_a=v_a/c$ is its relativistic beta factor with velocity $v_a$ and speed of light $c$, $\alpha$ is the fine structure constant, and $m_e$ is the mass of the electron. The axioelectric coupling $g_{ae}$ is the parameter on which we set a limit. Under the assumption of non-relativistic dark matter ALPs ($E_a=m_a c^2$ and $\beta_a\ll1$) constituting all galactic dark matter, the event rate for the axioelectric effect \cite{Arisaka:2012pb} can be expressed as a function of the axioelectric coupling $g_{ae}$:

\begin{equation}
    R_a  = \rho_{\chi}
\frac{3 g_{ae}^2 c }{16 \pi \,\alpha } \frac{m_a}{m_e^2} \sigma_{pe}(m_a c^2).
    \label{eqn:gae_rate}
\end{equation}

\subsubsection{Dark Photon Absorption}
The kinetic mixing of dark photons to SM photons enables an effective coupling to electrons, and with it the absorption of dark photons by atoms. The cross section for this process \cite{Bloch:2016sjj} is given by

\begin{equation}
    \sigma_{V}(E_V) =  \frac{\epsilon^2}{\beta_V}\sigma_{pe}(E_V),
\end{equation}

\noindent where the index $V$ denotes the dark photon, $\epsilon$ is the kinetic mixing parameter and $\beta_V=v_V/c$ is the dark photon's relativistic beta factor. Under the same assumptions as above (non-relativistic dark photons which constitute all of the galactic dark matter), the event rate as a function of the kinetic mixing parameter $\epsilon$ \cite{Bloch:2016sjj} is given by:

\begin{equation}
    R_V = \frac{\rho_{\chi}}{m_V} \epsilon^2 \sigma_{pe}(m_V c^2) c.
    \label{eqn:eps_rate}
\end{equation}

While in-medium effects can alter the effective kinetic mixing parameter that is probed through the dark absorption channel, this correction is only necessary for dark photon masses $\lesssim20$\,eV/$c^2$ \cite{Hochberg:2016sqx}, which is below the mass range considered in this analysis.

\subsubsection{\label{sec:PE} Photoelectric Cross Section}
\begin{figure}[!htb]
\centering
\includegraphics[width=1.0\columnwidth]{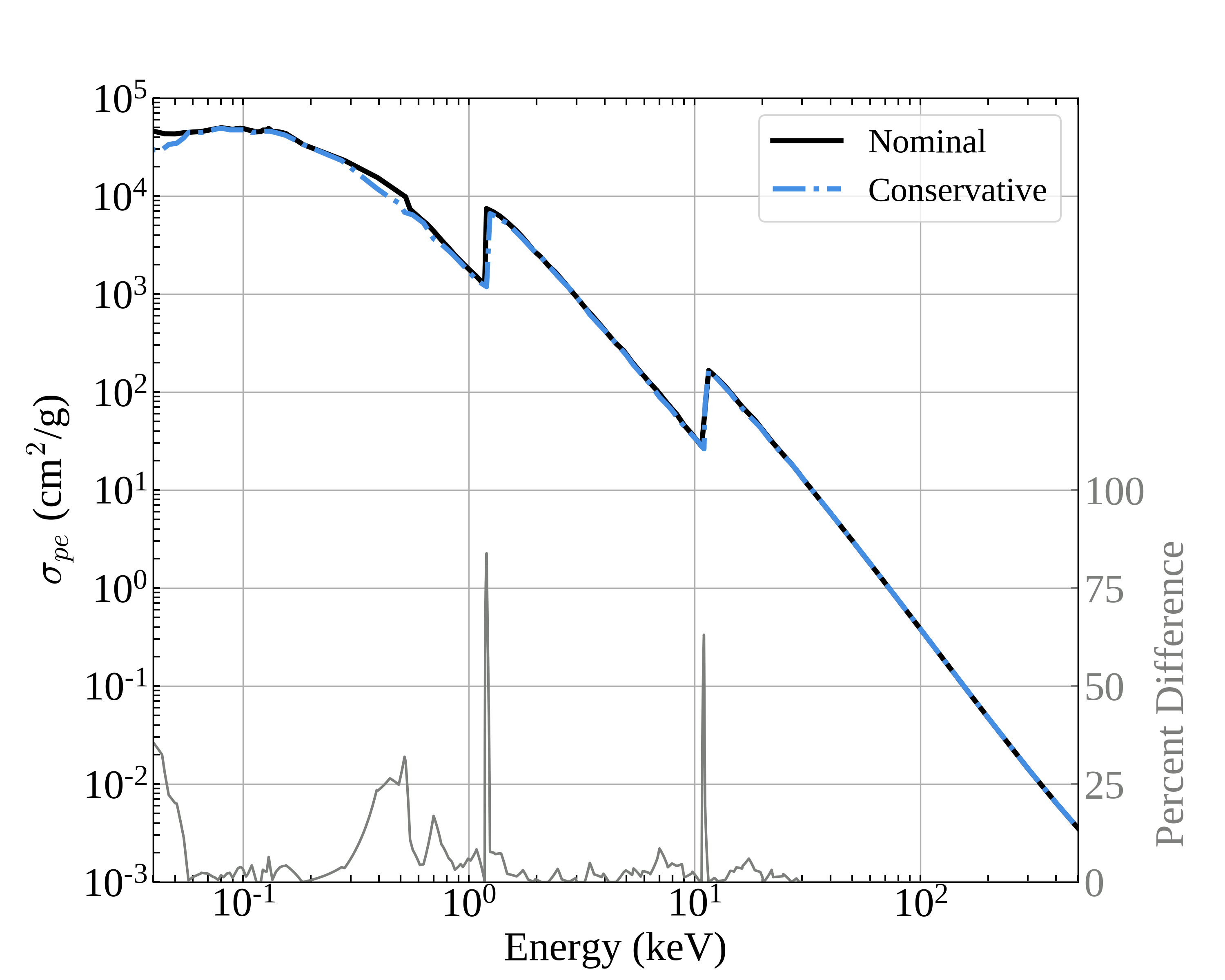}
\caption{
Photoelectric cross section $\sigma_{pe}$ for Ge as a function of energy in units of cm$^2$/g. The use of these macroscopic units means that Eqs.\,\ref{eqn:gae_rate} and \ref{eqn:eps_rate} are the interaction rates per target mass. The solid black line is the nominal cross section, using data from Refs.\,\cite{HENKE1993181,POTTER1997465,XCOM}.  The dashed blue line is the conservative cross section, determined by extracting the smallest $\sigma_{pe}$ values found in literature. The grey line at the bottom of the figure shows the percent difference between the nominal and conservative photoelectric cross sections. The two sharp peaks near the atomic binding energies are artifacts of the finite size of the energy steps in the data found in the literature search.
}\label{fig:xsex_ge}. 
\end{figure}

The cross sections of interest depend on the photoelectric cross section $\sigma_{pe}$ of the target, Ge. The $\sigma_{pe}$ data in the analysis range of 40\,eV to 500\,keV are expected to be approximately independent of temperature \cite{Hochberg:2016sqx}. Discrepancies were found in the literature \cite{POTTER1997465,HENKE1993181,XCOM,PhysRev.160.602,ProceedingsTenthInternational,zna-1973-0507,SovPhysSS,eScholarship,Hunter} across the analysis range, mostly for energies below 1\,keV. For this reason we use both a nominal and a conservative photoelectric cross section curve; the nominal curve is used to calculate our main results while we use the conservative one in the estimate of the associated uncertainty. The construction of the nominal $\sigma_{pe}$ curve follows the approach taken in Ref.\,\cite{Hochberg:2016sqx}, with data from Refs.\,\cite{POTTER1997465}, \cite{HENKE1993181}, and \cite{XCOM} for photon energies below 1\,keV, from (1--20)\,keV, and above 20\,keV respectively. The conservative $\sigma_{pe}$ curve was determined by using the smallest $\sigma_{pe}$ values found in the literature search, which results in the largest (most conservative) implied values of $\epsilon$ or $g_{ae}$ for a given measured rate. As much of the data found in the literature search were presented in plots, the data were extracted using a digitization web tool \cite{Webplotdigitizer}. Both the nominal and the conservative photoelectric cross sections are shown in Fig.\,\ref{fig:xsex_ge}. 

\section{\label{sec:cdms} SuperCDMS Soudan Setup}


SuperCDMS searches for dark matter interactions in cryogenic semiconductor detectors. The SuperCDMS Soudan detectors consisted of cylindrical Ge crystals, 25\,mm thick with a diameter of 76\,mm and a mass of $\sim$0.6\,kg. The high-purity crystals were instrumented on their top and bottom faces with hundred of tungsten transition edge sensors (TES). The TES arrays were interleaved with biased electrodes, used to collect charge carriers liberated by particle interactions in the detector substrates. The detectors were grouped in 5 stacks (towers, T1 to T5) with 3 detectors each (labeled Z1 through Z3).


The detector towers were located in the innermost of a set of nested copper cans for thermal shielding, surrounded by layers of polyethylene and lead shielding against environmental radioactivity, and a layer of scintillator panels to identify and discard interactions caused by residual cosmogenic radiation; see Ref.\,\cite{PRD72.2005.052009} for details. 

The total measured phonon energy from both primary recoil phonons and NTL phonons (the additional phonons produced by charges moving through the detector in the presence of an applied electric field) is given by

\begin{equation}
\label{eqn:energy}
E_t = E_r \left(1+\frac{eV_{\mathrm{bias}}}{\varepsilon_{eh}} \right),    
\end{equation}

\noindent where $E_r$ is the primary recoil energy, $e$ is magnitude of the electron charge, $V_{\mathrm{bias}}$ is the applied bias voltage and $\varepsilon_{eh}$ is the average energy required to produce an electron-hole pair ($\varepsilon_{eh}$ = 3.0\,eV for electron-recoil interactions in Ge \cite{PhysRev.140.A2089,Pehl}). In turn, the recoil energy can be expressed in terms of the measured phonon energy $E_t$ and the measured charge signal $E_q$ as

\begin{equation}
\label{eqn:recoilenergy}
E_r = E_t - E_q\frac{eV_{\mathrm{bias}}}{\varepsilon_{eh}}
\end{equation}

For iZIP detectors a 4\,V bias voltage across the detector was used. The sensor layout of the iZIP detectors \cite{Agnese:2013} made it possible to identify and discard events near the top and bottom surfaces, where signals could suffer from reduced charge collection efficiency. 

The data used for the iZIP analysis were taken between October 2012 and July 2013, from four of the original seven detectors included in the low-mass WIMP search, described in Ref. \cite{LT}. The reasons for this selection are described in Sec.\,\ref{sec:izip}. 

For CDMSlite \cite{PhysRevLett.112.041302}, the charge and phonon sensors on one side of a detector were set to a voltage bias of $\sim$70\,V while the sensors on the opposite side were grounded; only the phonon sensors on the grounded side were read out. For the chosen bias voltage, an amplification factor of more than 20 is achieved for the phonon energy from electron-recoil events. The large intrinsic amplification causes the measured signal to saturate for energies exceeding approximately 25--30\,keV. 

Beginning in 2012, SuperCDMS Soudan operated individual detectors in CDMSlite mode. In total, three temporally separated data sets, referred to as runs, were acquired: Run 1 was a proof of principle with a single detector \cite{PhysRevLett.112.041302}, Run 2 used the same detector for an extended period of time to yield an improved dark matter search result \cite{Agnese:2015nto}, and Run 3 --- with slightly less exposure than Run 2 --- was performed with a different detector \cite{Agnese:2018gze}. During Run 3, a change in operating conditions caused the phonon noise performance to worsen, motivating the decision to separate the analysis of Run 3 into two parts, referred to as Period 1 and Period 2.

Data from Run 2 \cite{Agnese:2015nto} and Run 3 \cite{Agnese:2018gze} were used for the CDMSlite part of the analysis discussed here. CDMSlite Run 2 has a lower threshold, a larger exposure and a moderately lower background than Run 3 (see Ref. \cite{Agnese:2018gze} for a discussion of the difference in backgrounds between the two runs). 
The main limitation for this analysis is the background, which leads to a similar sensitivity for both runs.

Interspersed throughout the dark matter search, calibration data were taken using $^{133}$Ba and $^{252}$Cf sources. Neutrons from the $^{252}$Cf source 
activated the detectors, producing $^{71}$Ge that decays via electron capture with a half-life of 11.43 days \cite{PhysRevC.31.666}. The resulting K-, L-, and M-capture lines at 10.37~keV, 1.30~kev and 160~eV are used for energy calibrations, as are the gamma absorption lines from the $^{133}$Ba source.

More details of the experimental setup, the different operating modes, and past analyses for SuperCDMS Soudan can be found in Refs.\,\cite{PRD72.2005.052009,PhysRevLett.120.061802,PhysRevLett.112.041302,LT,Agnese:2017jvy,Agnese:2018gze}.

\section{Analysis}
\label{sec:analysis}

To relate the observed event rate in the region of interest to an interaction rate, the detection and event selection efficiencies must be determined. In addition, for dark absorption of non-relativistic particles, the primary signal is a fixed energy deposition; the expected signature in the measured energy spectrum is a Gaussian peak at the energy corresponding to the candidate particle's mass, with a width given by the resolution of the detector. Therefore, it is also necessary to characterize each detector's energy resolution. 


The dark matter mass range under consideration for the CDMSlite data is 40\,~eV$/c^2$ to 25\,keV$/c^2$, where the lower limit is motivated by the limit setting method (see Sec. \ref{sec:limit}). In the iZIP analysis the considered masses range from 3 to 500\,keV$/c^2$, where the lower limit is chosen to avoid the rapidly dropping efficiency at lower energies. For CDMSlite we use the same selection criteria used for the Run 2 and Run 3 WIMP searches, for which the resolution model and detection efficiencies are already published \cite{Agnese:2015nto, Agnese:2018gze, Agnese:2018kta}.  Sec.\,\ref{sec:cdmslite} summarizes these results. Electron recoils in iZIP detectors have not previously been the focus of a dark matter analysis. As such, a reanalysis of iZIP electron recoils was necessary. Section \,\ref{sec:izip} describes the details of the iZIP analysis, including event selection criteria and their efficiencies, the energy scale calibration, and the resolution model.

In Sec.\,\ref{sec:limit} we motivate the definition of the analysis window size, the selection method for detectors to be included in the analysis (for a particular dark matter mass), and the technique used to combine data from different detectors to produce an upper limit on the rate.

\subsection{\label{sec:cdmslite}CDMSlite}

CDMSlite Run 2 data were acquired in 2014 between February and November. CDMSlite Run 3 Period 1 and Period 2 data were acquired in 2015 from February to the end of March and from April to May, respectively. The total detector exposures --- defined as the product of detector live time and mass --- are shown in Table\,\ref{tbl:cdmslite_exp} for each run. 

\begin{table}[ht]
    \centering
        \caption{\label{tbl:cdmslite_exp} CDMSlite exposures for Run 2 and Run 3 Periods 1 and 2.}
    \begin{tabularx}{\columnwidth}{XXXX}
        \hline
        \hline
         & Run 2 & Run 3-1 & Run 3-2\\
        \hline
        Exposure (kg$\cdot$days) & \multirow{2}{*}{70.1} & \multirow{2}{*}{31.5} & \multirow{2}{*}{29.4} \\
        \hline 
        \hline
    \end{tabularx}
\end{table}

\subsubsection{Event Selection and Signal Efficiency}
\label{sec:lite_eff}
\begin{figure}[!htb]
    \centering
    \includegraphics[width=1.0\columnwidth]{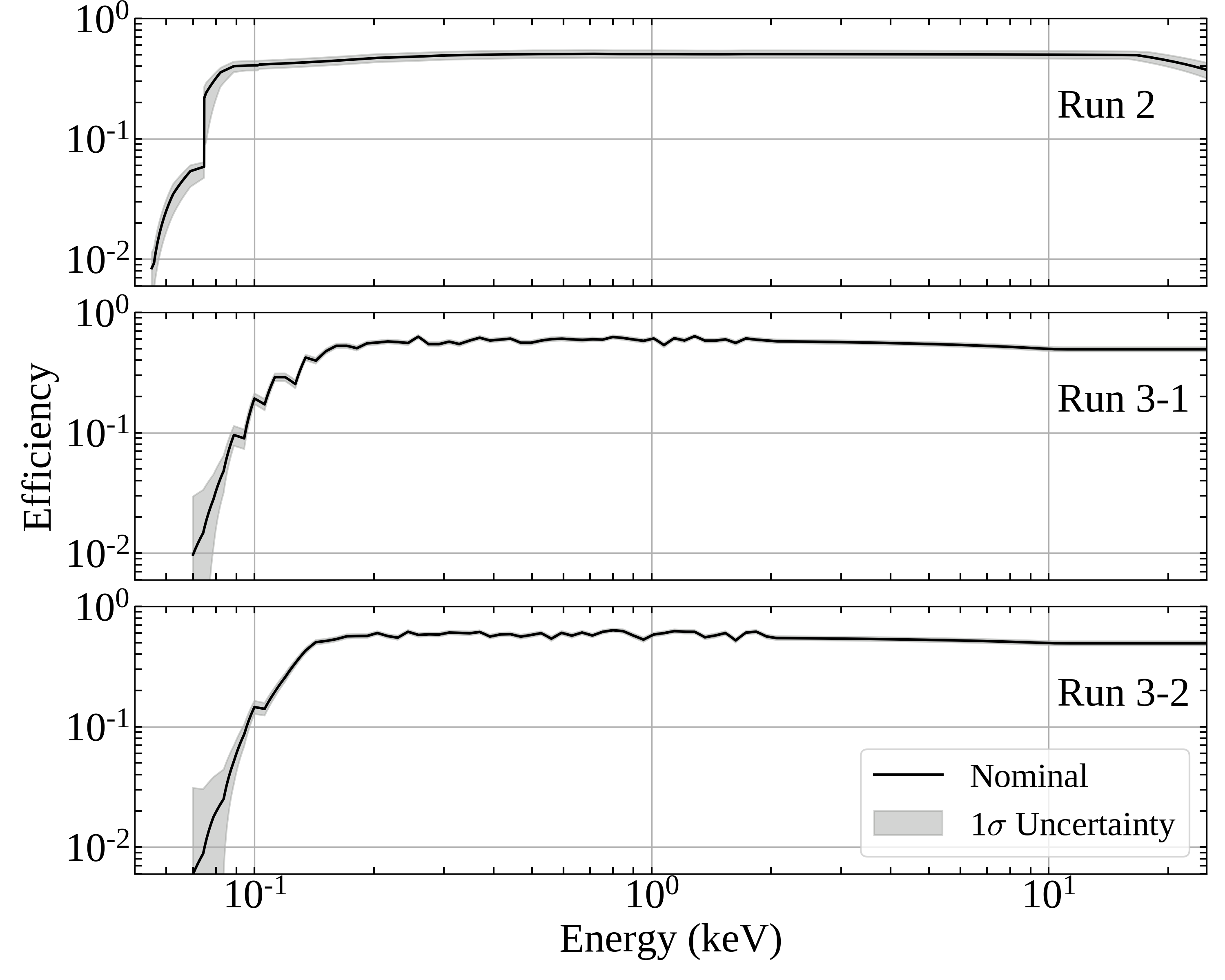}
    \caption{\label{fig:CDMSliteEfficiency}Signal efficiency for CDMSlite Run 2 \cite{Agnese:2017jvy}, Run 3 Period 1, and Run 3 Period 2 in the top, middle, and bottom subplots respectively. The nominal efficiencies (solid lines) and $1\sigma$ uncertainty bands (shaded regions) are shown for the full analysis range.}
\end{figure}

The event selection criteria, and therefore the signal efficiency and associated uncertainties, are the same as those used for the Run 2 and Run 3 WIMP searches \cite{Agnese:2015nto, Agnese:2018gze}. The signal efficiency shown in Fig.\,\ref{fig:CDMSliteEfficiency} describes the fraction of all recorded detector events that met all of the data selection criteria at a particular measured energy.

At low energy the efficiency is mostly determined by the trigger \cite{Bauer:2011}. For Run~2, a trigger threshold as low as 56\,eV was reached. In Run~3, the trigger rate at low energies was dominated by noise-induced events. These events were removed in the analysis based on their pulse shape, which lowered the efficiency and raised the effective threshold 
to 70\,eV. For events above approximately 100\,eV in Run 2 and 200\,eV in Run 3, the reduction in efficiency has little energy dependence and largely results from the radial fiducialization which removes events near the edge of the detector where inhomogeneities in the electric field lead to reduced NTL amplification. For a detailed discussion of the systematic and statistical uncertainties on the efficiency, see Refs.\,\cite{Agnese:2017jvy,Agnese:2018gze}. 

The original WIMP search analyses of CDMSlite Run 2 and 3 extended up to 2\,keV and the efficiency curves that were only derived up to this energy do not cover the full range needed for this analysis. For CDMSlite Run 2, the efficiency curve was extended up to 30\,keV for a background study \cite{Agnese:2018kta}. This was accomplished by linearly interpolating the efficiency between the values measured at 2\,keV and the $^{71}$Ge peak at 10.37\,keV; above this energy, electron recoils from $^{133}$Ba calibration data were compared to Monte Carlo simulations, showing a drop in efficiency starting at $\sim$20\,keV, which is attributed to saturation in one of the phonon sensors (see Ref.\,\cite{Agnese:2018kta} for details). For this analysis, the efficiency for Run 3 was extended in the same manner. However, the saturation effect observed in Run 2 occurs at a higher energy in the Run 3 detector, leading to a constant efficiency above the Ge K-shell line and below the upper analysis threshold. For both runs, the uncertainty on the efficiency is extended beyond 2\,keV in the same manner as the efficiency itself.

\subsubsection{Resolution Model}

We use the resolution model developed in the original CDMSlite analyses \cite{Agnese:2017jvy, Agnese:2018gze}, given by

\begin{equation}\label{eqn:res_mod}
    \sigma_T = \sqrt{\sigma_E^2 + \sigma_F^2(E) + \sigma_{PD}^2(E)},
\end{equation}

\noindent where $E$ is the measured energy, $\sigma_E$ is the baseline noise resolution, $\sigma_F$ describes the impact of the electron-hole pair statistics (accounting for the Fano factor \cite{Fano:1947zz}), and $\sigma_{PD}$ contributes a term to the resolution that is linear in energy and accounts for factors such as position dependence in the detector. The latter two quantities are energy dependent and are parameterized as $\sigma_F =\sqrt{BE}$ and $\sigma_{PD} = AE$, where $A$ and $B$ are constants. The resulting three free parameters of the resolution model, $A$, $B$ and $\sigma_E$, are constrained by the observed resolution of the easily identifiable K-, L-, and M-shell electron capture peaks of Ge, as well as the baseline width that results from electronic noise. The resolution of each of these peaks is extracted from Gaussian fits; Eq.\,\ref{eqn:res_mod} is then fit to the widths of these Gaussians (weighted by their uncertainties) at their respective energies, to extract the model parameters and their uncertainties from the fit. These are listed in Table\,\ref{tbl:cdmslite_res}. 

\begin{figure}[!htb]
    \centering
    \includegraphics[width=1.0\columnwidth]{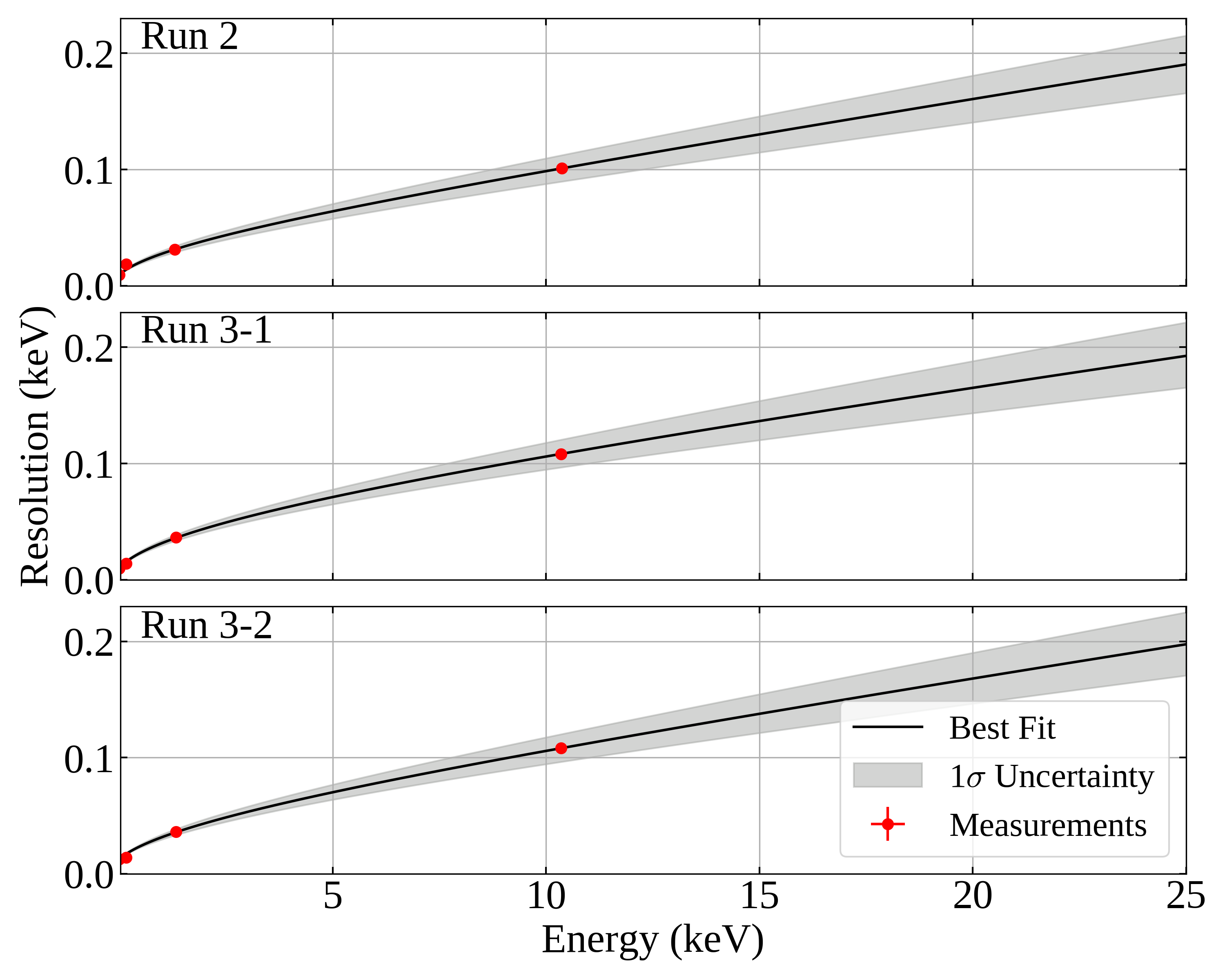}
    \caption{\label{fig:CDMSliteResolution}The energy resolution models for CDMSlite Run 2 \cite{Agnese:2017jvy} and  Run 3 \cite{Agnese:2018gze}. The best-fit curves (solid lines) and $1\sigma$ uncertainty bands (shaded regions) are shown with the measured widths (points, with $1\sigma$ error bars too small to see on this scale) of the three Ge electron capture peaks and the baseline noise resolution. The top, middle, and bottom panels show the models for Run 2, Run 3 Period 1, and Run 3 Period 2, respectively.}
\end{figure}

The resulting resolution model is plotted in Fig.\,\ref{fig:CDMSliteResolution}, separately for CDMSlite Run 2, Run 3 Period 1, and Run 3 Period 2. While the highest energy point used in the fits is the Ge K-shell capture line at 10.37\,keV, we assume that the fitted model is accurate for the entire analysis range. More details on the resolution model and a discussion of the uncertainties on the fit parameters can be found in Refs. \cite{Agnese:2017jvy, Agnese:2018gze}. The upper (lower) uncertainty band is formed by evaluating the resolution model with the best-fit parameters plus (minus) their uncertainties. The resulting $1\sigma$ uncertainty band is more conservative than the published bands in Refs.\,\cite{Agnese:2017jvy, Agnese:2018gze}.

\begin{table}[ht]
    \centering
    \caption{\label{tbl:cdmslite_res}Parameters for the resolution model with uncertainties for CDMSlite Run 2 and Run 3 Periods 1 and 2.}

    \begin{tabularx}{\columnwidth}{XXXX}
    \hline
    \hline
        & $\sigma_E$ (eV) & $B$ (eV) & $A$ ($\times$~$10^{-3})$ \\
        \hline
        Run 2 & 9.26 $\pm$ 0.11 & 0.64 $\pm$ 0.11 & 5.68 $\pm$ 0.94\\ 
        Run 3-1 & 9.87 $\pm$ 0.04 & 0.87 $\pm$ 0.12 & 4.94 $\pm$ 1.27\\ 
        Run 3-2 & 12.7 $\pm$ 0.04 & 0.80 $\pm$ 0.12 & 5.49 $\pm$ 1.13\\ 
    \hline
    \hline
    \end{tabularx}
\end{table}

\subsection{iZIP}
\label{sec:izip}
 The iZIP portion of this analysis draws data from the SuperCDMS low-mass WIMP search in Ref. \cite{LT}, acquired with seven detectors between Oct 2012 and July 2013 \cite{LT}. The same live-time selection is used here, which excludes periods directly following the $^{252}$Cf calibrations. Here we exclude data from two of the seven detectors due to shorts on one or more readout channels (phonon and charge), and we exclude data from a third detector that shows evidence of incomplete charge collection on one side. The four detectors selected for this analysis are listed in Table\,\ref{tab:iZIPdetectors} along with their exposures. 

\begin{table}[ht]
    \centering
        \caption{\label{tab:iZIPdetectors}The four selected iZIPs and their exposures.}
    \begin{tabularx}{\columnwidth}{XXXXX}
        \hline
        \hline
         & T1Z1 & T2Z1 & T2Z2 & T4Z3 \\
        \hline
        Exposure (kg$\cdot$days) & \multirow{2}{*}{80.2} & \multirow{2}{*}{82.9} & \multirow{2}{*}{80.9} & \multirow{2}{*}{83.8}\\
        \hline
        \hline
    \end{tabularx}
\end{table}

\subsubsection{Event Selection and Signal Efficiency}
\label{sec:zip_eff}

To select events, we require that they fulfill a set of data quality criteria (data quality cut), that the detector in question issued a trigger (trigger requirement) while no other detector did (single-scatter cut), that the event was not coincident with a signal in the muon veto detector (muon veto cut), and that they pass a fiducial volume cut. The first four criteria are identical to those described in Ref.\,\cite{LT} and have a combined efficiency of around 95\,\% for each detector, with slight energy dependence near the analysis threshold. 

The role of the fiducial volume cut is to reject the $^{210}$Pb surface-event background and to remove events near the edge and close to the flat surfaces of the detector where the distorted electric field may lead to a reduced NTL gain. The fiducial volume cut developed for nuclear recoils, and its efficiency, was determined using $^{252}$Cf calibration data in the low-mass WIMP search of Ref. \cite{LT}. This cut definition uses the charge signals, which differ significantly between nuclear and electron recoils, and its efficiency is energy dependent. As the signal of interest here consists of single-scatter (and thus essentially point-like) electron-recoil events, the signal efficiency should be re-evaluated using this type of events. Most electron recoils originate from gamma interactions which often scatter multiple times within the detector and thus cannot be used as proxy for the signal events. However, there is one identifiable sample of single scatter events in our data set: the Ge K-shell capture events at $E_K=10.37$\,keV. Since these events only appear at a fixed energy, they cannot be used to measure the energy dependence of the efficiency of this cut. Therefore, a new fiducial volume cut was developed (based exclusively on the charge signal distribution and therefore referred to as the ``charge fiducial volume" or QFV cut) with the goal to make its efficiency largely energy independent so that it can be measured using the Ge K-shell events. This cut consists of two components, a radial charge cut to remove events near the cylindrical surface and a charge symmetry cut to remove events near the flat surfaces. Below we describe each of these cuts and how we assess the energy dependence of their efficiency, before we discuss how the overall efficiency of the combined cut is determined.\\

\begin{figure}[!htb]
    \centering
    \includegraphics[width=1.0\columnwidth]{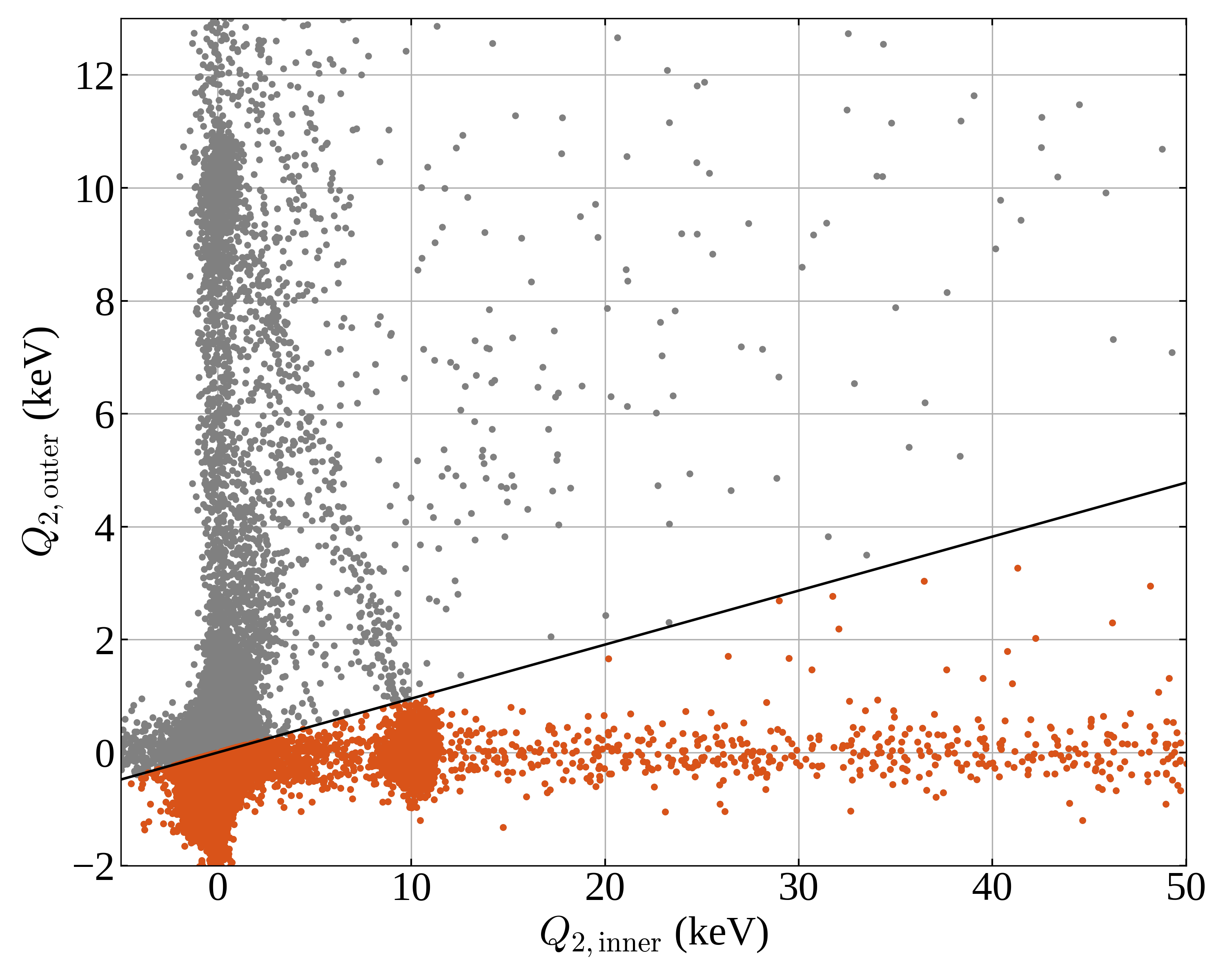}
    \caption{\label{fig:T2Z2_rad_cut}Signal amplitude in the outer vs.\ the inner charge channel for side 2 of detector T2Z2 ($Q_{2, \mathrm{outer}}$ vs $Q_{2, \mathrm{inner}}$). The cut line that removes events with an outer signal above a given fraction of the total signal is shown. At energies below $E_K$ (10.37~keV) the line cuts into the noise distribution, removing some of the events with a real outer fraction below the threshold. This effect is modelled and accounted for in the final efficiency calculation.}
\end{figure}

\paragraph{Radial Charge Cut} The radial charge cut removes events where the charge collected in the outer electrodes exceeds a certain fraction of the total charge collected\footnote{Note that this definition differs from that of the radial charge cut developed in Ref.\,\cite{LT}, where events with any discernible signal in the outer electrodes were removed.}. This fraction is determined at the Ge K-shell peak in the total charge spectrum (inner plus outer electrode). The signal in the outer electrode is required to be less than three times the baseline resolution of that electrode, as measured using the events that appear at $E_K$ in the inner electrode. This new definition of the radial cut fulfills the requirement of a constant efficiency above the K-shell energy, since the energy distribution between inner and outer signal is energy-independent for point-like events occurring at a given position in the detector. Due to the noise in the measurement, events at lower energy with very little or no charge collected on the outer electrode still have a significant chance to have a reconstructed amplitude above the cut limit and thus fail the cut
(see Fig.\,\ref{fig:T2Z2_rad_cut}). 
This effect is quantified by modeling the signal distribution between the inner and outer charge electrodes using the Ge K-shell events. This model is then scaled to lower energies and convolved with the applicable charge resolution before using it to determine the radial charge cut efficiency.

\begin{figure}[!htb]
    \centering
\includegraphics[width=1.0\columnwidth]{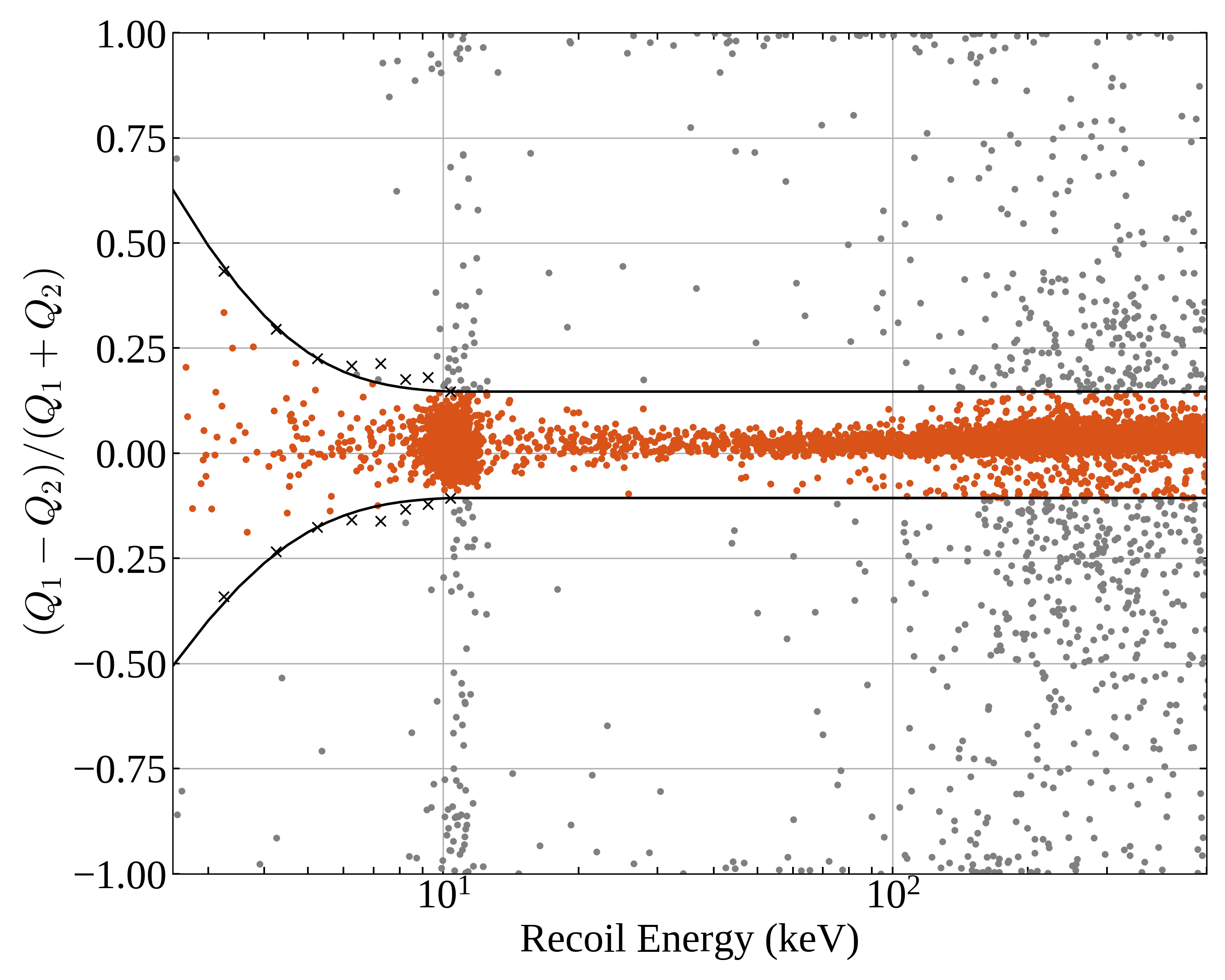}
    \caption{\label{fig:T2Z2_sym_cut}Charge symmetry cut parameter vs.\ total charge signal for detector T2Z2, where $Q_1$ is the total charge collected on side 1, and $Q_2$ is the total charge collected on side 2. Bulk events are localized in a narrow band around zero that widens towards low energy. A Gaussian distribution is fit to the band in 1\,keV bins at low energy and the position of three times the standard deviation is indicated by black crosses in the figure. The cut line is an exponential function fit to these points below $E_K$ with a smooth transition to a horizontal line above $E_K$. The distribution continues to narrow above $E_K$, so a constant cut value together with the assumption of a constant efficiency (as function of energy) will lead to a conservative upper limit on the extracted rate, compensating for any uncertainty that is introduced by the higher density of events near the cut line at high energies.}
\end{figure}

\paragraph{Charge Symmetry Cut \label{sec:izip_sym}} Events occurring in the bulk of the detector are expected to have symmetric charge collection on the sensors on both sides of the detector, while events near one of the flat surfaces have reduced or no signal on the opposite side. Events near the top and bottom surfaces can thus be removed by requiring that the signal amplitude is similar between the two sides (``charge symmetry"). In Ref. \cite{LT} the charge symmetry cut was defined using only the inner charge signals on both sides of the detector. With our new definition of the radial charge cut, the total charge signal on each side of the detector may include a contribution from the outer electrode. Therefore, the charge symmetry cut is redefined for this analysis based on the total charge signal on each side. A cut parameter is defined as the ratio of the difference between the charge collected on each side to the sum of the total charge collected on both sides. For detector-bulk events at a given energy, the distribution of this parameter is roughly Gaussian, centered near zero. The distribution is widest at low energy, before narrowing to an approximately constant width at energies above $\sim$10\,keV. From 1\,keV to the K-shell electron capture peak energy, the mean $\mu$ and width $\sigma$ of the distribution are measured in 1\,keV energy bins, and events outside of $\mu\pm3\sigma$ are removed. In the range above $E_K$ the cut is set to stay constant at the value determined at $E_K$ (see Fig.\,\ref{fig:T2Z2_sym_cut}).

Below $E_K$, the total number of events just outside the cut boundary is very small, so the final spectrum does not change significantly even if the cut position is loosened far beyond possible uncertainties. This guarantees that the assumption of a constant efficiency below $E_K$ is conservative. At higher energies, the event density near the bulk distribution is higher overall so that the exact choice of the cut position has a greater influence on the efficiency, and small non-Gaussianities of the distribution become more relevant. The constant cut value, together with the still slightly narrowing distribution suggests a moderately increasing efficiency. However, in absence of a method to measure the precise efficiency value, we assume a constant efficiency which leads to a conservative limit in the final analysis.

\paragraph{QFV Cut Efficiency at $E_K$}
The final step is to determine the efficiency of the new QFV cut for the (single-scatter) Ge K-shell events by exploiting the short 11.43~day half-life of $^{71}$Ge \cite{PhysRevC.31.666}. For this we use the data with a particularly high $^{71}$Ge decay rate acquired over 10 or 20 day periods directly after the $^{252}$Cf neutron calibrations that were excluded from the dark matter search.  The livetime, data quality, and veto criteria are identical to those used in the dark matter analysis. The approach used is similar to that used in \cite{Agnese:2015nto}. Events with recoil energy within a $\pm4\sigma$ window of the Ge K-shell electron capture peak are selected, where $\sigma$ is the energy resolution at $E_K$. The recoil energy is determined using Eq.\,\ref{eqn:recoilenergy} to ensure that we select all true K-shell events events, including those with reduced NTL gain that are removed by the QFV cuts. The selected events are divided into two categories: (1) those passing the QFV cut and with total phonon energy within the signal region ($E_K\pm1\sigma$, see Sec.\,\ref{sec:limit} below for a discussion on the choice of the signal region), and (2) those failing the former criteria. 

The number of K-shell events in each category is extracted via a likelihood fit to the time distribution of events, where time is measured from the most recent $^{252}$Cf calibration period. The model probability distribution function $P(t,r)$ is the sum of an exponential (decaying with the $^{71}$Ge lifetime $\tau$) and a constant background component:

\begin{equation}
    P(t,r) = r\,e^{-t/\tau} + (1-r),
\end{equation}

\noindent where $t$ is the time since the last $^{252}$Cf calibration period and $r$ is the ratio between the number of $^{71}$Ge events that are represented by the exponential component and all other events that are represented by the flat component. The efficiency is then the number of $^{71}$Ge events in the signal region divided by the sum of the $^{71}$Ge events in both categories. The uncertainty on the the fit result is determined by calculating the likelihood as a function of $r$ and extracting the $1\sigma$ confidence interval from the resulting distribution. Note that this efficiency is the combined efficiency of the QFV cut and the signal window selection. This ranges from $30-36$\,\%, depending on the detector. 

\begin{figure}[!htb]
    \centering
    \includegraphics[width=1.0\columnwidth]{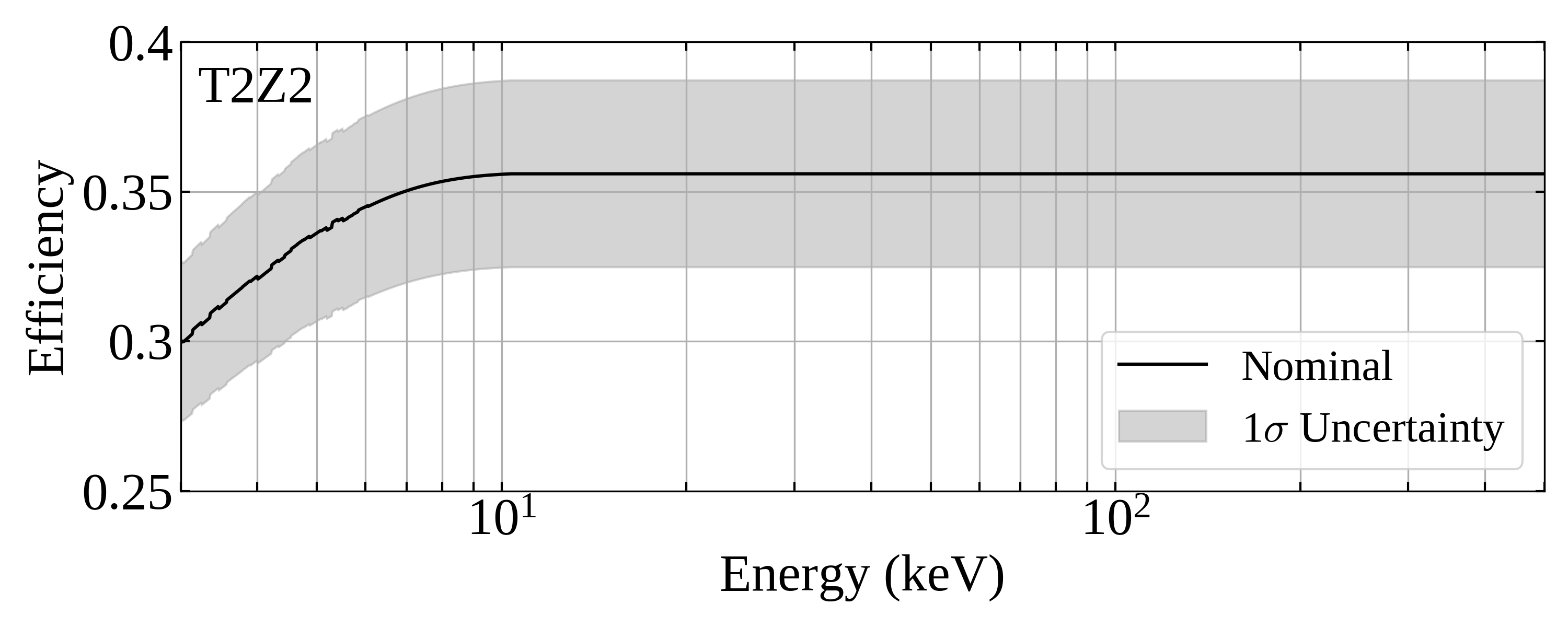}
    \caption{\label{fig:T2Z2_eff}The efficiency for detector T2Z2 used in the present iZIP analysis. The nominal value (solid line) and $1\sigma$ uncertainty (shaded region) are shown.}
\end{figure}

\paragraph{Combined Efficiency} The combined efficiency of all selection criteria is constant above $\sim$5\,keV for all selected detectors; the uncertainty on this (approximately 3\,\% for all detectors) is dominated by the statistical uncertainty on the QFV efficiency from the likelihood fit. As an example, the total combined efficiency for detector T2Z2 is shown in Fig.\,\ref{fig:T2Z2_eff}; all detectors exhibit similar behaviour.   

\subsubsection{Energy Calibration and Resolution Model}
With the selection criteria applied, the events with reduced NTL amplification are removed, and we can calibrate the energy scale for all events using a simple scaling function for the conversion between the measured total phonon energy $E_t$ and the recoil energy $E_r$ instead of applying an event-by-event correction using Eq.\,\ref{eqn:energy}. A quadratic function is fit to the ratio of measured phonon signal to true energy for peaks that are used for the iZIP calibration: the Ge K-shell peak at 10.37\,keV, the 66.7\,keV peak that appears in $^{252}$Cf calibration data as consequence of inelastic neutron scattering, the 356\,keV peak from $^{133}$Ba calibration data, and the 511\,keV peak from positron annihilation that is observed in dark matter search data. The L- and M-shell peaks used for fitting the resolution model in the CDMSlite analysis are below the analysis threshold of 3\,keV imposed for the iZIPs in this analysis (see Sec.\,\ref{sec:limit} below). Also, the baseline noise peak is not used, as at the true energy of 0\,keV the ratio of measured to true energy is undefined. Defining the energy scale in this manner accounts for detector saturation at higher energies.

\begin{figure}[!htb]
    \centering
    \includegraphics[width=1.0\columnwidth]{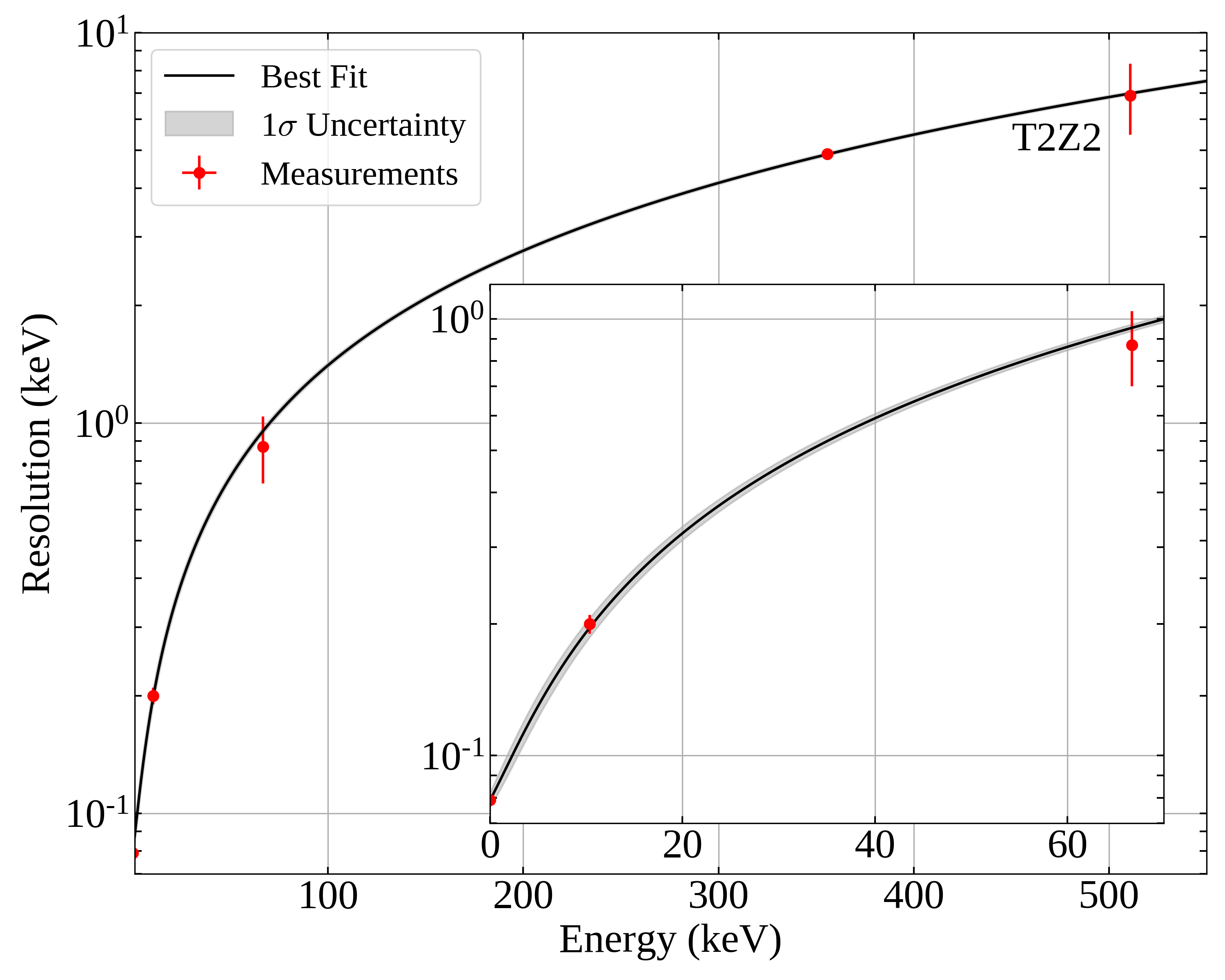}
    \caption{\label{fig:T2Z2_res}The fitted resolution model used in the iZIP analysis, for the example of detector T2Z2. The measured peak widths (points) are used to determine the best-fit curve (solid line) and $1\sigma$ uncertainty band (shaded region). The inset plot shows the same data, zoomed in to the region between 0 and 70\,keV.}
\end{figure}

The functional form of the resolution model used for the iZIPs is the same as the one used for CDMSlite (see Eq.\,\ref{eqn:res_mod}). The model is fit to the resolution of five peaks weighted by their uncertainties: the four peaks used for calibrating the energy scale and the baseline noise peak. The model parameters and the 1$\sigma$ uncertainty band on the resolution are determined for each detector individually. The upper (lower) edge of the band is formed by taking the upper (lower) value of the $1\sigma$ confidence interval for each parameter determined by the fit. The resulting model parameters and their statistical uncertainties from the fit (systematic uncertainties are comparatively negligible) are shown in Table\,\ref{tbl:izip_res}. The model for T2Z2, including the measured peak energies and widths to which the model is fit, is shown as an example in Fig.\,\ref{fig:T2Z2_res}. 

\begin{table}[ht]
    \centering
        \caption{\label{tbl:izip_res}Resolution model parameters and their  uncertainties for the iZIP detectors included in this analysis.}
    \begin{tabularx}{\columnwidth}{XXXX}
    \hline
    \hline
         & $\sigma_E$ (eV) & $B$ (eV) & $A$ ($\times$~$10^{-3})$ \\
        \hline
        T1Z1 & 100.5 $\pm$ 3.4 & 2.7 $\pm$ 0.5 & 19.5 $\pm$ 0.1\\ 
        T2Z1 & 69.9 $\pm$ 4.6 & 2.4 $\pm$ 1.1 & 15.4 $\pm$ 0.3\\ 
        T2Z2 & 79.0 $\pm$ 2.9 & 1.2 $\pm$ 0.3 & 13.6 $\pm$ 0.1\\ 
        T4Z3 & 80.2 $\pm$ 5.4 & 0.6 $\pm$ 1.2 & 19.0 $\pm$ 0.4\\ 
    \hline
    \hline
    \end{tabularx}
\end{table}

\subsection{Limit Setting}
\label{sec:limit}
An upper limit on the rate for the three CDMSlite and four iZIP data sets. This section describes the calculation of the limit on each data set, and the method by which the limits are combined. No background modeling is performed in this analysis, therefore, only an upper limit on the signal strength can be extracted. The signal model, a Gaussian centered at the dark matter mass with the width determined by the detector's resolution at that energy, is the same for the interactions of ALPs and dark photons. To set a limit on the axioelectric coupling and the dark photon kinetic mixing parameter, we first set a limit on the observed rate of signal events. Limits on the physical quantities of interest are then extracted using Eq.\,\ref{eqn:gae_rate} and Eq.\,\ref{eqn:eps_rate}, for $g_{ae}$ and $\epsilon$ respectively. To determine a limit on the interaction rate for a particular dark matter mass, the events in a window around the mass equivalent energy in the spectrum are counted and a Poisson upper limit on that number is calculated, which is then divided by the efficiency-corrected exposure to convert it to a limit on the rate. The process is repeated for each data set with its corresponding event spectrum, resolution model, and efficiency, before the different data sets are combined.

The choice of window size is a compromise between maximizing the signal efficiency (large window) and minimizing the included background (small window). The optimal window size depends on the observed event numbers. A window size of of $\pm1\sigma$ was selected which is close to the optimum for numbers on the order of one to a few tens of events per $\sigma$ (where $\sigma$ is the resolution at the respective mass-equivalent energy), which are the typical values found in our data sets.

However, the event selection efficiency drops quickly near the trigger threshold and the expected signal shape (the Gaussian times the efficiency function) is no longer centered about the mass-equivalent energy, but skewed towards higher energies. In this case, setting the cut at $+1\sigma$ may remove the dominant fraction of the recorded signal. Therefore, for energies close to the trigger threshold, the upper limit of the window is chosen based on the expected signal shape rather than the primary Gaussian shape. The cut value is the $+1\sigma$-equivalent, cutting the same 15.9\% that $+1\sigma$ would cut from a Gaussian. The lower edge of the window is kept at $-1\sigma$. This change in window choice is implemented for energy depositions in the CDMSlite detector below 100\,eV for Run 2 and below 200\,eV for Run 3. 

Dark matter masses below the trigger threshold can be studied due to the positive tail of the expected event distribution. Since the efficiency estimate for masses far below the trigger threshold is strongly impacted by potential non-Gaussian tails and the uncertainty on the resolution, the lower bound of the CDMSlite analysis range is chosen to be roughly $2\sigma$ below the trigger threshold.

For the iZIP detectors we keep the $\pm1\sigma$ window throughout, but we limit the dark matter analysis to an energy range above 3\,keV, which is well above the trigger threshold (around 1\,keV) so that the discussed effect due to rapidly decreasing signal efficiency is small. Since the best sensitivity at low masses is expected to be derived from CDMSlite data anyway (due to the superior energy resolution), this imposed threshold is not expected to limit the sensitivity of this analysis.

The upper edge of the dark matter analysis range of 25\,keV for the CDMSlite data sets is set to avoid saturation effects caused by the large intrinsic signal amplification. For the iZIP detectors, the upper limit of 500\,keV is just below the energy of the highest available peak (511\,keV) used for calibration and to measure the resolution. 

Limits on the event rate are calculated for a set of narrowly spaced discrete dark matter masses. The limits are calculated separately for each data set which includes the energy range corresponding to each given dark matter mass. 

Differences in background rates between detectors call for a method that enables the calculation of a combined limit that is not unnecessarily weakened by data sets with background rates that are truly higher, while still minimizing selection bias from statistical fluctuations when only including low-rate data sets. 
 
For a given dark matter mass the following procedure is implemented:
\begin{enumerate}
\item Select the data set with the lowest observed rate.
\item Discard any data set where the difference between its observed rate and the lowest rate exceeds three times the uncertainty on this difference (3$\sigma$).
\item Include any remaining data sets in the combined analysis.
\end{enumerate}

Given the small number of data sets, a $3\sigma$ deviation in the rates will rarely occur if all detectors truly measure the same rate and differences are only due to statistical fluctuations. However, discarding data sets still may introduce a small bias, so a Monte Carlo simulation was performed to quantify this bias. For the case that all detectors indeed measure the same rate, the bias was found to be less than one percent, given the typical observed rates and the number of detectors involved. Since the actual rate is not the same in all detectors, the true bias is lower, and can thus be considered negligible.

The measured event numbers of all included data sets (within the energy window for a particular dark matter mass) are summed and the statistical 90\,\% upper limit on that number is calculated. This upper limit is then divided by the combined efficiency-weighted exposure to determine the 90\,\% upper limit on the measured event rate. The actual number of data sets that are included for a particular dark matter mass varies as a function of energy due to variation of background contributions in each detector; in some energy ranges only one detector is kept, whereas for other ranges all detectors are included.

With the bias of the selection of data sets to be included in the analysis shown to be negligible, the main remaining systematic uncertainty in the data analysis as described here is the uncertainty on the efficiency of the charge symmetry cut for the iZIP detectors. As discussed above, the uncertainty of the charge symmetry cut efficiency is negligible at low energy compared to the uncertainties already explicitly included in the uncertainty band on the efficiency curve (particularly the statistical uncertainty from the method of determining the efficiency at the Ge K-shell peak, see Section \ref{sec:zip_eff}); at high energy, the choice made in the assumption of the efficiency accounts for possible systematics and leads to a conservative limit. In the CDMSlite analysis the dominant uncertainty is also the uncertainty on the efficiency, where the main contributing factor is again the fiducial volume cut. The uncertainty on the energy resolution is subdominant. Further details of the uncertainties in the CDMSlite analysis can be found in \cite{Agnese:2015nto,Agnese:2018gze}. We also assessed the impact of the assumption that single-scatter events are point-like. Using data from the NIST online data base ESTAR \cite{ESTAR}, we estimated the effect of a finite extension of the highest energy events considered in the analysis to be less than 2\,\% and thus negligible compared to other uncertainties. So, for the final uncertainty on the rate limit, we only include the uncertainty bands on the resolution and efficiency curves as shown in the respective figures above. 

For a given mass, the upper (lower) bound for the rate limit is generally determined using the upper (lower) limit on the resolution and the lower (upper) limit on the efficiency. However, the alternative method of determining the analysis window for low masses in CDMSlite can lead to a situation where a wider resolution leads to a better sensitivity. For these masses the upper and lower uncertainty bands are determined by calculating the limits on the rate for all combinations of the upper and lower limits of the resolution and efficiency, and picking the lowest and highest rate limits, respectively.

Three combined rate limits were produced:
(1) a combined iZIP limit from the four iZIP detectors, (2) a combined CDMSlite limit, and, in the energy range from 3-25\,keV/$c^2$, where both CDMSlite and iZIP detectors are used, (3) an overall combined limit using all seven data sets (CDMSlite Run 2, Run 3 Periods 1 and 2, and four iZIP detectors). The calculated rate limits and their uncertainties are shown in Fig.\,\ref{fig:rates}. 

\begin{figure}[!htb]
    \centering
    \includegraphics[width=1.0\columnwidth]{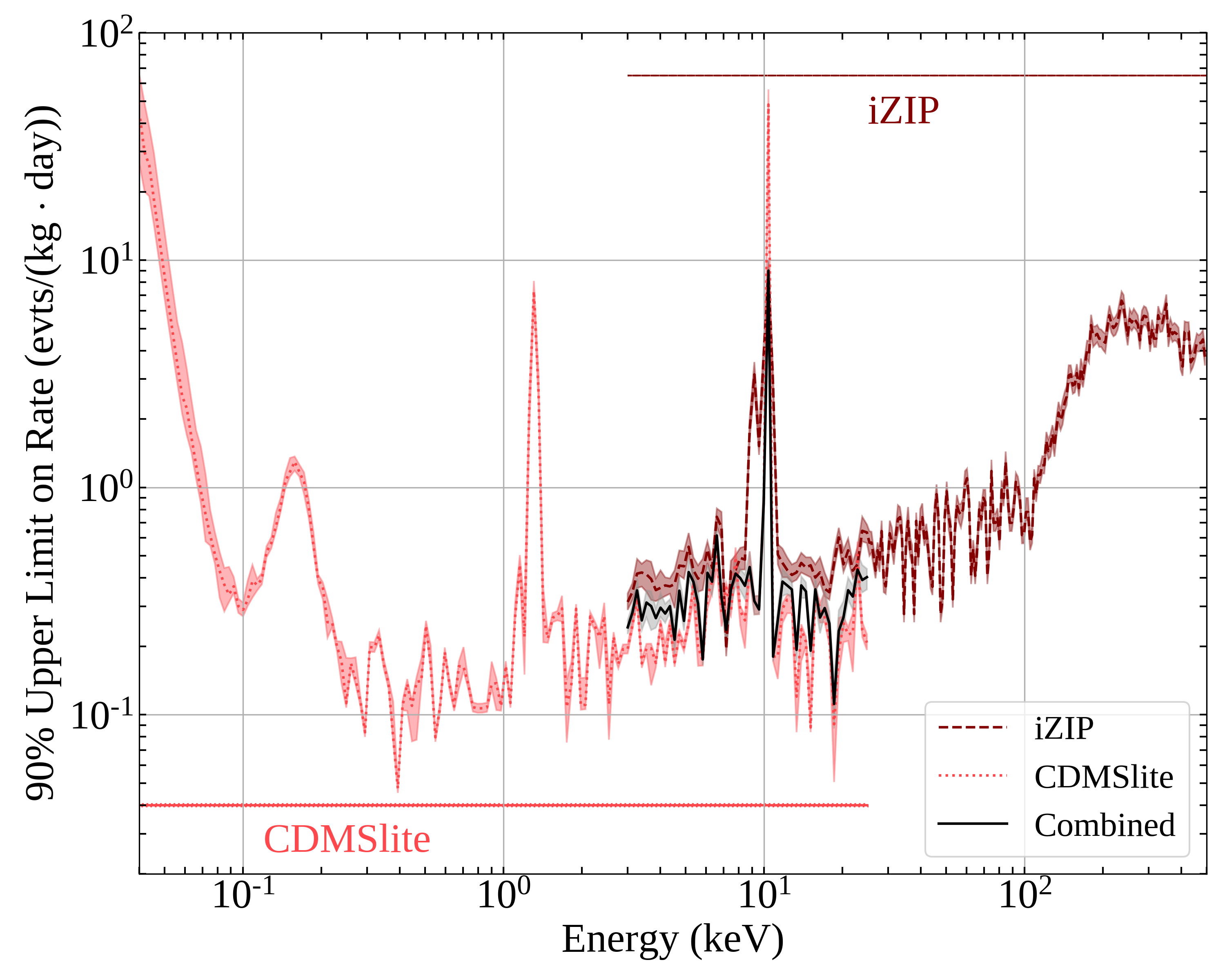}
    \caption{\label{fig:rates} Calculated rates from combining data from the CDMSlite runs (light red dotted line), the iZIP detectors (dark red dashed line), and all data sets (black solid line). The shaded region around each line corresponds to the uncertainty on the rate limits. The elevation in the limits near 10\,keV, 1\,keV, 160\,eV and in iZIP data around 9\,keV results from the higher measured rates due to the Ge K-, L- and M-captures and the K-capture of $^{65}$Zn respectively.}
\end{figure}

In the overlap between the CDMSlite and iZIP analysis range ($3-25$\,keV/$c^2$), the CDMSlite rate limit is generally stronger than the iZIP limit. This is expected if background rates are comparable, since in CDMSlite mode the detectors have a better energy resolution (about a factor of 2 at 10\,keV, compare Fig.\,\ref{fig:CDMSliteResolution} and Fig.\,\ref{fig:T2Z2_res}). Additionally, the iZIP data analyzed here were acquired during the first measurement period (late 2012 to mid 2013), while the CDMSlite data were acquired towards the end of SuperCDMS Soudan (early 2014 - mid 2015); therefore, some of the cosmogenic activity (e.g.\ $^{65}$Zn at 9\,keV or $^{55}$Fe around 6\,keV) has decayed to a significantly lower rate. However, the difference in rate is small enough that the method of selecting detectors for the combined limit still includes iZIP detectors, so the combined limit is typically between that from CDMSlite and iZIPs. There is a data point near 7\,keV/$c^2$ where the combined limit is stronger than the iZIP or CDMSlite limits separately. Here the rate of all detectors included in the combination is similar and the combined limit benefits from the improved statistics. 

\section{\label{sec:results}Results}

The limits on the rate are converted to limits on the relevant physical quantities, $g_{ae}$ and $\epsilon$, using Eq.\,\ref{eqn:gae_rate} and Eq.\,\ref{eqn:eps_rate}, respectively. The upper (lower) bounds of the uncertainty bands are calculated by combining the upper (lower) bound on the limit on the rate with the conservative (nominal) photoelectric cross section. 

An additional systematic uncertainty stems from the measurement of $\rho_{\chi}$, which enters linearly in the calculation of the final limits. However, the results in literature to which we compare our limits (see below) use the same assumption of $\rho_{\chi}=$~0.3\,GeV\,/($c^{2} \,\mathrm{cm}^{3})$; thus we chose not to reflect this uncertainty in our final results.

The results on the search for ALPs and dark photons are shown in Fig.\,\ref{fig:gae_lim} and Fig.\,\ref{fig:eps_lim} in the form of exclusion limits on the axioelectric coupling $g_{ae}$ and the dark photon kinetic mixing parameter $\epsilon$, respectively. Limits from other direct-detection experiments and astrophysical constraints from models of stellar cooling are also shown for comparison. 

\begin{figure}[!htb]
    \centering
    \includegraphics[width=1.0\columnwidth]{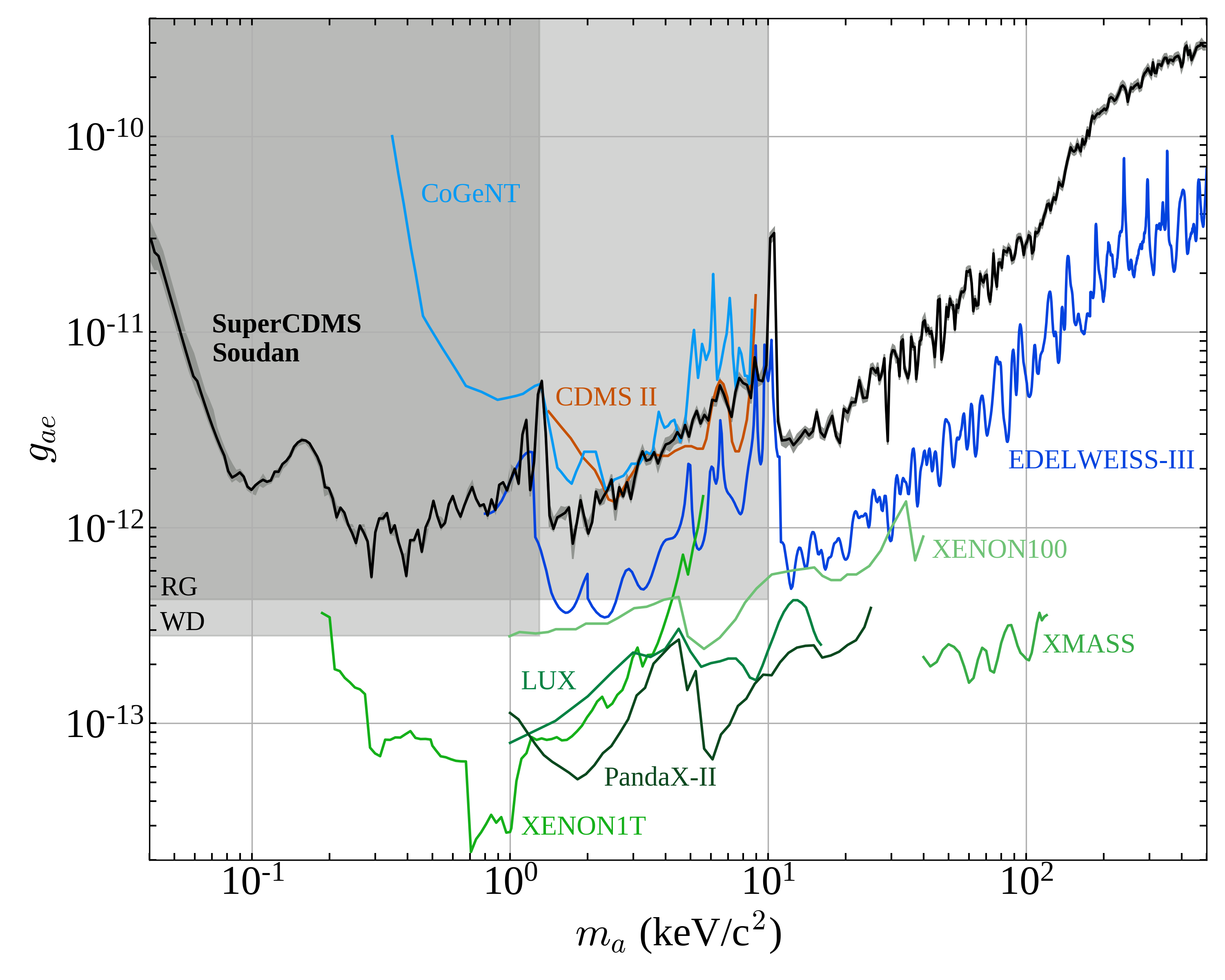}
    \caption{\label{fig:gae_lim} SuperCDMS Soudan upper limit (solid black) with uncertainty band (shaded grey) on the axioelectric coupling. Also shown are limits set by other direct-detection experiments including: CDMS II \cite{PhysRevLett.103.141802}, CoGeNT \cite{PhysRevLett.101.251301}, EDELWEISS-III \cite{Armengaud:2018cuy}, LUX \cite{Akerib:2017uem}, PandaX-II \cite{Fu:2017lfc}, XENON100 \cite{Aprile:2014eoa}, XENON1T \cite{Aprile:2019xxb}, and XMASS \cite{Abe:2018owy}. The shaded regions are excluded by the observed cooling of red giant (RG) \cite{PhysRevLett.111.231301, PhysRevD.98.030001} and white dwarf (WD) stars \cite{Bloch:2016sjj, PhysRevD.98.030001}.}
\end{figure}

\begin{figure}[!htb]
    \centering
    \includegraphics[width=1.0\columnwidth]{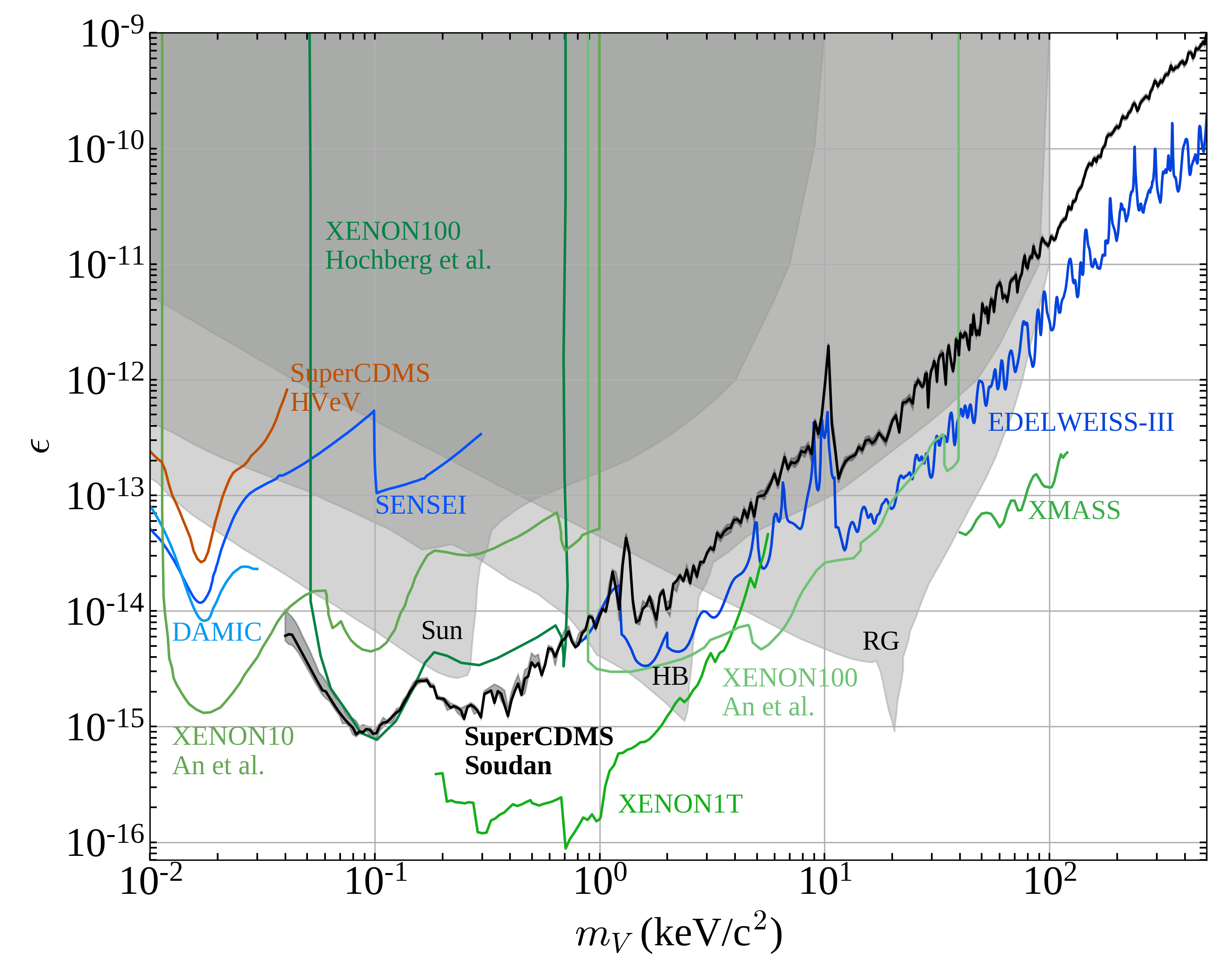}
    \caption{\label{fig:eps_lim} SuperCDMS Soudan upper limit (solid black) with uncertainty band (shaded grey) on the dark photon's kinetic mixing. Also shown are limits set by other direct-detection experiments including: DAMIC \cite{Aguilar-Arevalo:2019wdi}, EDELWEISS-III \cite{Armengaud:2018cuy}, SENSEI \cite{Abramoff:2019dfb}, the SuperCDMS HVeV device \cite{Agnese:2018col}, An \textit{et al}.'s analysis of XENON10 and XENON100 \cite{Pospelov:2015}, Hochberg \textit{et al}.'s analysis of XENON100 \cite{Hochberg:2016sqx}, XENON1T \cite{Aprile:2019xxb}, and XMASS \cite{Abe:2018owy}. The shaded regions show limits set from anomalous energy loss mechanisms in the Sun, horizontal branch stars (HB), and red giants (RG) from \cite{Pospelov:2015}. }
\end{figure}

\section{\label{sec:conclusion}Conclusion}
This analysis sets the strongest laboratory constraint on galactic dark matter ALPs in the mass range $(40-186)$\,eV/$c^2$. The absolute values of the laboratory limits depend on the assumption that the respective species constitutes all of the galactic dark matter with a local density of 0.3\,GeV\,/($c^{2} \mathrm{cm}^{3}$). Astrophysical constraints on the observed cooling of white dwarf \cite{Bloch:2016sjj, PhysRevD.98.030001} and red giant \cite{PhysRevLett.111.231301, PhysRevD.98.030001} stars set the strongest exclusion limits below 1\,keV. It should be noted though that a different set of assumptions are used to produce the astrophysical constraints. For example, the production rates of ALPs in a stellar environment requires a precise understanding of the energy levels and occupation levels for each nucleus in the star. In practice this is done with state of the art models of the radiative opacities in the Sun \cite{Redondo:2013wwa}.

World leading or competitive limits are also set on the kinetic mixing of dark photons in the mass range of (40-186)\,eV/$c^2$. Astrophysical limits \cite{Pospelov:2015} from horizontal branch stars, red giants, and the Sun are weaker than those from direct detection experiments below 1\,keV/$c^2$. \\

\section*{ACKNOWLEDGMENTS}
The  SuperCDMS  collaboration  gratefully  acknowledges  technical  assistance  from  the staff  of  the  Soudan Underground Laboratory and the Minnesota Department of  Natural  Resources. The  iZIP  detectors  were  fabricated in the Stanford Nanofabrication Facility, which is a member of the National Nanofabrication Infrastructure Network,  sponsored  and  supported  by  the  NSF.  Funding  and  support  were  received  from  the  National  Science  Foundation,  the  U.S. Department  of  Energy,  Fermilab URA Visiting Scholar Grant No.\ 15-S-33, NSERC Canada, the Canada Excellence  Research  Chair  Fund, the McDonald Institute (Canada), MultiDark (Spanish MINECO), and the Deutsche Forschungsgemeinschaft (DFG, German Research Foundation) - Project No.\,420484612. The SuperCDMS collaboration prepared this document using the resources of the Fermi National Accelerator Laboratory (Fermilab), a  U.S.  Department  of Energy, Office  of  Science,  HEP User  Facility. Fermilab  is  managed  by  Fermi  Research Alliance,  LLC (FRA), acting  under  Contract  No. DE-AC02-07CH11359. Pacific  Northwest  National  Laboratory  is  operated by  Battelle  Memorial  Institute  under Contract No. DE-AC05-76RL01830 for the U.S. Department of Energy. SLAC is operated under Contract No. DEAC02-76SF00515  with  the  U.S. Department  of  Energy.

\clearpage
\bibliography{main}

\clearpage

\onecolumngrid

\begin{center}

{\large \bf Correction: Constraints on dark photons and axion-like particles from SuperCDMS Soudan} \\[0.25cm]

See above for author list

\end{center}

\vspace{0.5cm}

\twocolumngrid

In our publication describing the search for dark matter using electron recoil data from four iZIP and two CDMSlite detectors there is an error in the implementation of the limit setting method for the iZIP datasets.

The signal window is defined as $\pm1\sigma$ about the dark matter mass equivalent energy, where $\sigma$ is the resolution at that energy. The Poisson upper 90\% confidence limit on the number of events in the signal window was converted to an upper limit on the event rate by dividing by the efficiency-weighted exposure. The efficiency factor is calculated by integrating the product of the signal model and the energy-dependent efficiency over the signal window. For the iZIP datasets, the average efficiency over the signal window was mistakenly used, ignoring the effect that $\sim$32\% of the signal is lost due to the limited window size. The efficiency factors were therefore too large by a factor of $\sim$1.47. 

This error has no impact on the limits in the original publication for dark matter masses below 3$\,$ keV$/c^2$ and therefore does not affect the region in which we claim to set the strongest laboratory constraints.

\begin{figure}[H]
    \centering
    \includegraphics[width=1.0\columnwidth]{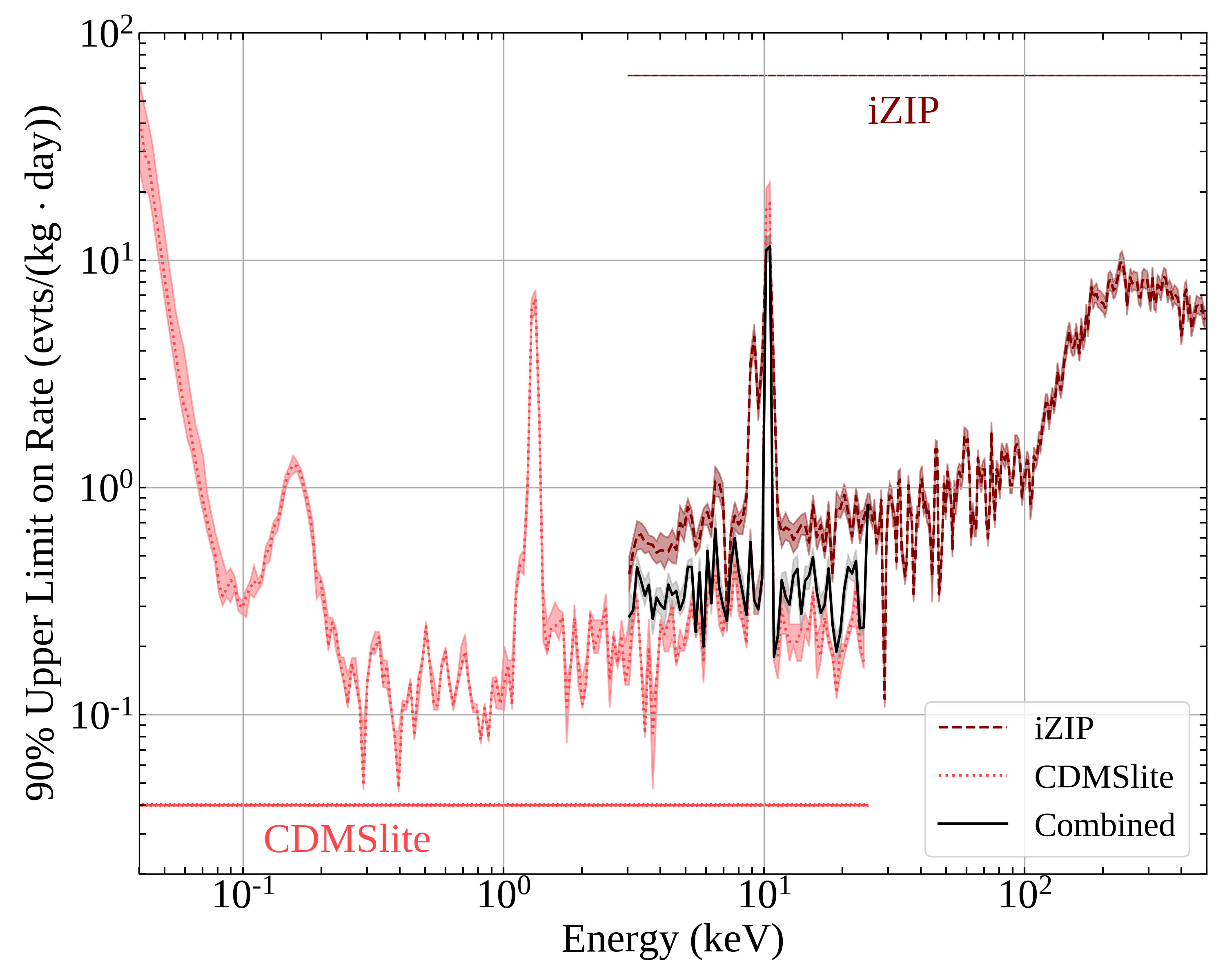}
    \caption{\label{fig:rates} Corrected calculated rates from combining data from the CDMSlite runs (light red dotted line), the iZIP detectors (dark red dashed line), and all data sets (black solid line). The shaded region around each line corresponds to the uncertainty on the rate limits. The elevation in the limits near 10\,keV, 1\,keV, 160\,eV and in iZIP data around 9\,keV results from the higher measured rates due to the Ge K-, L- and M-captures and the K-capture of $^{65}$Zn respectively.}
\end{figure}

\ \ 
\vspace{3cm}

Additionally, in Figures 8, 9, and 10 of the original publication, there is a minor inconsistency (but not inaccuracy) in the visualization, as the precise mass values for which the limits are shown vary slightly between the figures. The limits are calculated for a dense set of masses; for better clarity in the figures, the limit-curves are down-sampled by a factor of 50, but unintentionally a different subset of masses was picked for each figure. The three figures have now been made consistent, all using the masses shown in Figure 9 of the original publication.

Updated versions of Figures 8, 9, and 10 of the original publication are provided in this correction as Figures \ref{fig:rates}, \ref{fig:gae_lim}, and \ref{fig:eps_lim}, respectively. 

\vspace{2.6cm}

\begin{figure}[hbt!]
    \centering
    \includegraphics[width=1.0\columnwidth]{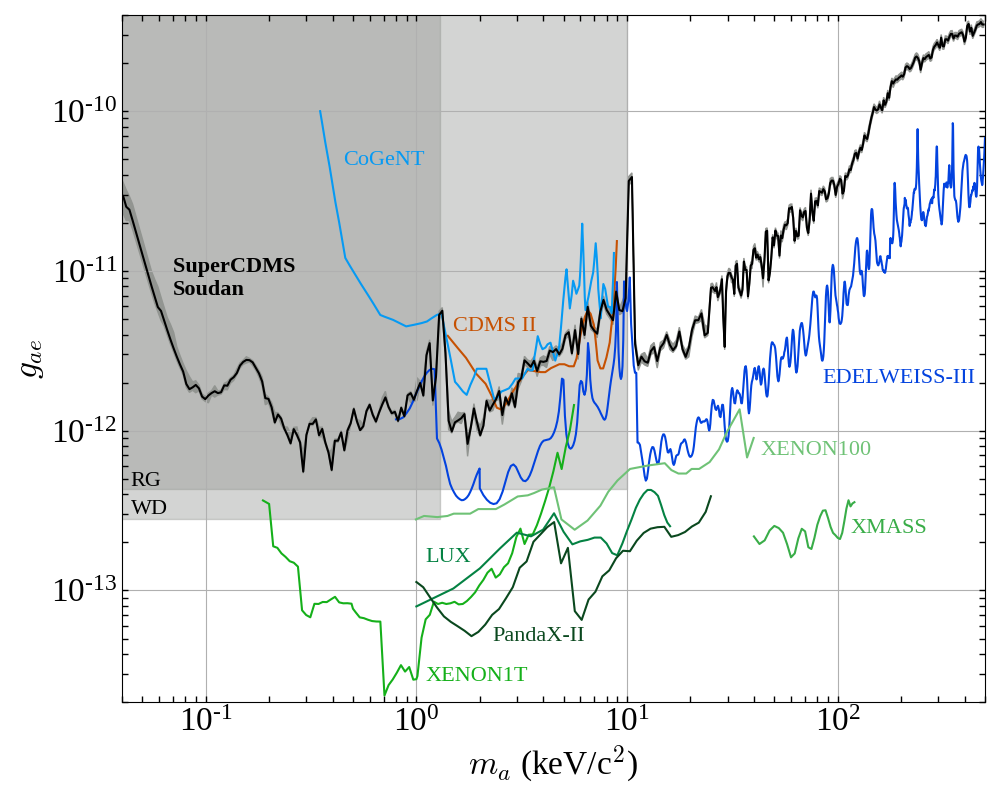}
    \caption{\label{fig:gae_lim} Corrected SuperCDMS Soudan upper limit (solid black) with uncertainty band (shaded grey) on the axioelectric coupling. Also shown are limits set by other direct-detection experiments including: CDMS II \cite{PhysRevLett.103.141802}, CoGeNT \cite{PhysRevLett.101.251301}, EDELWEISS-III \cite{Armengaud:2018cuy}, LUX \cite{Akerib:2017uem}, PandaX-II \cite{Fu:2017lfc}, XENON100 \cite{Aprile:2014eoa}, XENON1T \cite{Aprile:2019xxb}, and XMASS \cite{Abe:2018owy}. The shaded regions are excluded by the observed cooling of red giant (RG) \cite{PhysRevLett.111.231301, PhysRevD.98.030001} and white dwarf (WD) stars \cite{Bloch:2016sjj, PhysRevD.98.030001}.}
\end{figure}

\begin{figure}[h!]
    \centering
    \includegraphics[width=1.0\columnwidth]{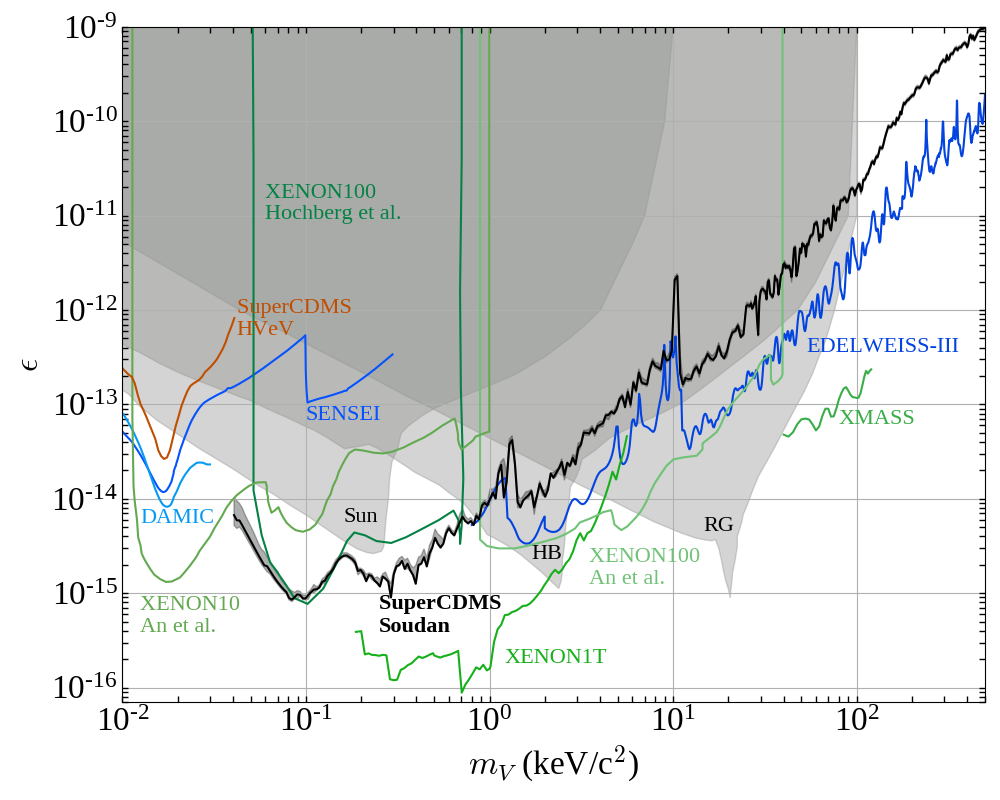}
    \caption{\label{fig:eps_lim} Corrected SuperCDMS Soudan upper limit (solid black) with uncertainty band (shaded grey) on the dark photon's kinetic mixing. Also shown are limits set by other direct-detection experiments including: DAMIC \cite{Aguilar-Arevalo:2019wdi}, EDELWEISS-III \cite{Armengaud:2018cuy}, SENSEI \cite{Abramoff:2019dfb}, the SuperCDMS HVeV device \cite{Agnese:2018col}, An \textit{et al}.'s analysis of XENON10 and XENON100 \cite{Pospelov:2015}, Hochberg \textit{et al}.'s analysis of XENON100 \cite{Hochberg:2016sqx}, XENON1T \cite{Aprile:2019xxb}, and XMASS \cite{Abe:2018owy}. The shaded regions show limits set from anomalous energy loss mechanisms in the Sun, horizontal branch stars (HB), and red giants (RG) from \cite{Pospelov:2015}. }
\end{figure}

\end{document}